\documentclass[12pt]{article}
\newcommand{\Comment}[1]{{}}
\usepackage{amsfonts,amsthm,amsmath,amssymb,slashed}
\usepackage[textwidth = 430 pt, textheight = 630 pt]{geometry}
\usepackage{bbold}

\Comment{\usepackage{color}
\definecolor{MyDarkBlue}{rgb}{0.15,0.15,0.45}
\usepackage[linktocpage=true]{hyperref}
\hypersetup{
colorlinks=true,
citecolor=MyDarkBlue,\bibliography{myref}
linkcolor=MyDarkBlue,
urlcolor=MyDarkBlue,
pdfauthor={Thiago Araujo and Horatiu Nastase},
pdftitle={. . .},
pdfsubject={hep-th}
}

\usepackage[numbers,sort&compress]{natbib}
\usepackage{hypernat}}
\usepackage{graphicx}

\newcommand\ignore[1]{}
\def\one{{\,\hbox{1\kern-.8mm l}}}

\newcommand{\tr}{\operatorname{tr}}

\def\Tr{{\rm Tr\, }}

\def\a{\alpha}\def\b{\beta}

\def\m{\mu}
\def\n{\nu}

\def\r{\rho}
\def\s{\sigma}

\def\l{\lambda}

\def\d{\partial}

\def\D{\Delta}

\newcommand{\Cset}{{\,\,{{{^{_{\pmb{\mid}}}}\kern-.45em{\mathrm C}}}}}

\newcommand{\dd}{\mathrm d}

\newcommand{\be}{\begin{equation}}
\newcommand{\bea}{\begin{eqnarray}}

\newcommand{\ee}{\end{equation}}
\newcommand{\eea}{\end{eqnarray}}

\usepackage{color}
\newcommand{\bse}{\begin{subequations}}
\newcommand{\ese}{\end{subequations}}
\usepackage{graphicx}
\usepackage{caption}
\usepackage{subcaption}
\usepackage{cite}
\newcommand{\bb}{\mathbb}

\begin{document}

\renewcommand{\thefootnote}{\fnsymbol{footnote}}

\makeatletter
\@addtoreset{equation}{section}
\makeatother
\renewcommand{\theequation}{\thesection.\arabic{equation}}

\rightline{}
\rightline{}
   \vspace{1.8truecm}


\vspace{10pt}


\begin{center}
{\LARGE \bf{\sc Observables in the Guarino-Jafferis-Varela/CS-SYM duality}}
\end{center}
 \vspace{1truecm}
\thispagestyle{empty} \centerline{
 {\large \bf {\sc Thiago R. Araujo${}^{a,b}$}}\footnote{E-mail address: \Comment{\href{mailto:thiago.araujo@apctp.org}}{\tt thiago.araujo@apctp.org}}
{\bf{\sc and}}
   {\large \bf {\sc Horatiu Nastase${}^{a,}$}}\footnote{E-mail address: \Comment{\href{mailto:nastase@ift.unesp.br}}{\tt
    nastase@ift.unesp.br}}
}

\vspace{1cm}

\vspace{.8cm}
\centerline{{\it ${}^a$
Instituto de F\'{i}sica Te\'{o}rica, UNESP-Universidade Estadual Paulista}} \centerline{{\it
R. Dr. Bento T. Ferraz 271, Bl. II, Sao Paulo 01140-070, SP, Brazil}}

\centerline{{\it ${}^b$
Asia Pacific Center for Theoretical Physics, Pohang 37673, Korea}} \centerline{{\it
Department of Physics, POSTECH, Pohang 37673, Korea}}

\vspace{1.0truecm}


\thispagestyle{empty}

\centerline{\sc Abstract}

\vspace{.4truecm}

\begin{center}
\begin{minipage}[c]{380pt}
{\noindent We study various semiclassical observables in the duality proposed by Guarino, Jafferis and Varela,
between a warped $AdS_4\times$ squashed $S^6$ gravitational solution and a 
3 dimensional ${\cal N}=2$ SYM-CS$_k$ conformal gauge theory, deformed from the maximal $SU(N)$ ${\cal N}=8$ SYM. 
Baryon vertices corresponding to particle-like branes have unusual behaviour with $N$ and $k$ and present strong evidence for 
a certain level-rank duality. Wilson loops and the anomalous dimensions of 
operators of high spin scale like $(N/k)^{3/2}$. The entanglement entropy behaves like in a usual CFT. Giant magnon operators obey the same 
law as in 4 dimensional ${\cal N}=4$ SYM, and giant gravitons are also sub-determinant operators. 

}
\end{minipage}
\end{center}

\today

\vspace{.5cm}

\setcounter{page}{0}
\setcounter{tocdepth}{2}

\newpage

\tableofcontents
\renewcommand{\thefootnote}{\arabic{footnote}}
\setcounter{footnote}{0}

\linespread{1.1}
\parskip 4pt

\newpage

\section{Introduction}

The AdS/CFT correspondence has proved very successful in obtaining the nonperturbative properties of strongly coupled field theories. The 
best understood is the one originally studied by Maldacena, the 
duality of 4 dimensional $SU(N)$ ${\cal N}=4$ SYM vs. string theory in $AdS_5\times S^5$, which has been used as a toy 
model for the physics of strong interactions, i.e. QCD \cite{Maldacena:1997re}. On the other hand, in 3 dimensions, the best understood duality is 
of the ABJM model vs. string theory in $AdS_4\times \mathbb{CP}^3$ background \cite{Aharony:2008ug}, which has been used as a toy model
for 3 dimensional condensed matter systems. The ABJM model is an ${\cal N}=6$ superconformal theory with gauge group $SU(N)\times SU(N)$ and 
Chern-Simons gauge fields, which is why it is more useful for condensed matter than for particle physics. In both SYM and ABJM cases a very large 
number of different types of calculations have been performed, due to the simplicity of the theories, yet in both the theories are very special (they are
toy models), so the applicability of the lessons learned is unclear. 

Recently however, a very interesting duality has been proposed in 3 dimensions by Guarino, Jafferis and Varela, between a fixed point of an ${\cal N}=2$
supersymmetric $SU(N)$ SYM gauge theory with a Chern-Simons term of level $k$ added and string theory in a certain 
warped, squashed $AdS_4\times S^6$ background of massive type IIA theory \cite{Guarino:2015jca}. The theory is still superconformal, but has 
less supersymmetry, has an extra parameter $k$ that forms a 't Hooft coupling $\lambda=N/k$, and has both the SYM term interesting for 
particle physics and a CS term at level $k$ that as we will see can dominate the IR and imply a certain interesting level-rank duality. 
The theory arises as a deformation of the maximal ${\cal N}=8$ $SU(N)$ SYM  (with adjoint fields) on the worldvolume of $N$ D2-branes
(see also \cite{Guarino:2016ynd}), and 
the resulting SCFT has been  studied in \cite{Schwarz:2004yj}.

The logic that led to the 
GJV solution was as follows. The four dimensional ${\cal N}=8$ $SO(8)$-gauged supergravity \cite{deWit:1982bul} 
arises as a consistent truncation of 11 dimensional supergravity (low energy of M-theory) on the seven sphere \cite{deWit:1986oxb, Nicolai:2011cy}. 
But it was pointed out in \cite{Dall'Agata:2012bb, Dall'Agata:2014ita, deWit:2013ija} that the original de Wit-Nicolai solution \cite{deWit:1982bul} is, in fact, 
just one particular point in a continuous one-parameter family of solutions, where the parameter $\omega\in [0,\pi/8]$ determines the linear combination 
of the $SO(8)$ electric and magnetic gauge fields. In this framework, the de Wit-Nicolai solution is purely electric and can be embedded into the 
$D=11$ supergravity as a consistent truncation. These dyonic gaugings cannot be embedded in string/M-theory, as recently proved in \cite{Lee:2015xga}.
A related symmetry group, namely the $ISO(7)=SO(7) \ltimes \mathbb{R}^7$,  can be obtained from $SO(8)$ through a In\"on\"u-Wigner contraction 
\cite{Hull:1984yy}. Solutions with this symmetry also admit dyonic gaugings \cite{Dall'Agata:2014ita}, studied further in \cite{Guarino:2015qaa},
and it has been proved \cite{Hull:1988jw} that 
the purely electric solutions can be embedded into $D=11$ supergravity as a consistent truncation on $S^6 \times S^1$. 
Moreover, the $ISO(7)$-dyonically gauged solution 
considered by GJV \cite{Guarino:2015jca} arises as a consistent truncation of massive type IIA solution on a squashed $6$-sphere (shown also 
in \cite{Guarino:2015vca}), 
that we denote by ${\cal S} S^5$, and preserves up to $\mathcal{N}=3$ supersymmetries 
\cite{Varela:2015uca, Guarino:2015jca, Pang:2015vna, Pang:2015rwd}. The magnetic coupling constant $m$ is identified with the Romans mass $F_0$. 
In fact, this solution has fixed points that uplift to ${\cal N}=2$ $AdS_4 \times {\cal S} S^5$.
Since the internal space has the topology of a sphere, we expect that this solution arises as the near horizon geometry of a stack of $N$ D$2$-branes, 
with the Romans mass inducing a Chern Simons term on the brane worldvolume.\footnote{A related duality conjecture has been considered in 
\cite{Fluder:2015eoa}, where the authors studied a stack of $D2$-branes probing a Calabi-Yau threefold singularity in the massive type IIA supergravity. 
In this case, the field theory is an ${\cal N}=2$ Chern-Simons quiver gauge theory $U(N)^G$, where $G$ is the Euler number of the resolved manifold. 
The authors in \cite{Fluder:2015eoa} argued that in the low energy, the D$2$-brane theory flows to a modification (resolution) of the 
gravity dual ${\cal N}=2$ $AdS_4 \times {\cal S} S^5$ of GJV \cite{Guarino:2015jca}. }

In this paper, we will study the GJV duality from the point of view of various observables associated with semiclassical string theory objects 
in the gravity dual. Static D-branes wrapped on cycles in the geometry correspond, if there are nontrivial tadpoles, to baryon vertices in the field theory, 
the (external) quarks corresponding to long strings that must end on the D-brane. Moving D-branes wrapped on cycles correspond to giant gravitons
(gravitons that polarize into nontrivial D-branes in the geometry), and correspond to certain subdeterminant operators. 
Long strings in the geometry can give Wilson loops, the anomalous dimension of high spin operators and giant magnon spin chain operators. 

The paper is organized as follows. In section two we review the GJV solution, perform the quantization of the charges, and review the dual 
field theory. In section 3 we study D-branes on cycles and their gravity duals: baryon operators for the static branes and determinant operators 
for moving branes, i.e. giant gravitons, as well as the gauge coupling on the wrapped branes. 
In section 4 we study entanglement entropy from the holographic prescription. In section 5 we study 
strings in the geometry, giving Wilson loops and the anomalous dimension of operators of large spin. In section 6 we study giant magnons from the 
point of view of both gravity and field theory, and in section 7 we conclude.


\section{Guarino-Jafferis-Varela solution and duality to SYM-CS}

In \cite{Guarino:2015jca} the authors showed that a consistent truncation of a massive type IIA SUGRA on the six sphere $S^6$ gives rise 
to the $ISO(7)$-dyonically-gauged SUGRA with the magnetic coupling constant related to the Romans mass. They also  showed that this 
supergravity solution has a fixed point which uplifts to a new ${\cal N}=2$ $AdS_4\times \mathcal{S} S^5$ solution, where 
$\mathcal{S} S^5$ is the suspension of $S^5$ and is topologically equivalent to the six sphere $S^6$ \cite{Fluder:2015eoa}.

The fixed point solution \cite{Guarino:2015jca, Fluder:2015eoa} has metric (in the Einstein frame and the conventions of \cite{Cvetic:1999un})
\bse
\be 
\dd s^2_{IIA}=\Delta\left\{ \dd s^2_{AdS_4}+\frac{3}{2}\dd \a^2 +\Xi  \dd s^2_{\mathbb{CP}^2}+\Omega\eta^2 \right\}\label{metric}
\ee
where $\D=L^2(3+\cos 2\a)^{1/2}(5+\cos 2\a)^{1/8}$, $\Xi=6\sin^2\a/(3+\cos2\a)$, $\Omega=9\sin^2\a/(5+\cos2\a)$ and $\a\in [0, \pi]$. Also 
$\eta=\dd \psi+\omega$, where $\psi\in [0,2\pi]$ and ${\cal J}=\frac{1}{2}\dd \omega$ is the K\"ahler-form of $\mathbb{CP}^2$, which is parameterized by the coordinates $(\l, \theta, \phi, \s)$, we refer to appendix A for further details. Finally, we write the warp factor in string frame as 
\be
L^2_{string}=e^{\phi/2}\Delta=L^2 e^{\phi_0/2}(5+\cos 2\a)^{1/2}.
\ee

The determinant of the internal manifold metric
\be 
\dd s^2_{{\cal S} S^5} = \frac{3}{2}\dd \a^2 +\Xi  \dd s^2_{\mathbb{CP}^2}+\Omega\eta^2 \label{int-space}
\ee
is
\be 
\det(G_{{{\cal S} S^5}})=\frac{3\  \Xi^4\ \Omega }{128} \sin^2\theta \sin^6\lambda \cos^2\lambda\; .\label{det}
\ee
For fixed $\psi$ and $\a$ (with $\a \neq 0, \pi$ to avoid the isolated conical singularities), the internal manifold has a $\mathbb{CP}^2$
space, and for $\l=\pi/2$ and fixed $\s$ it has a $\mathbb{CP}^1$.

The solution also contains the fields
\begin{align}
e^{\phi} & =e^{\phi_0}\frac{(5+\cos 2\a)^{3/4}}{(3+\cos 2\a)}\label{dilaton}\\
H_3&=24\sqrt{2}L^2 e^{\phi_0/2}\frac{\sin^3\a}{(3+\cos2\a)^2}{\cal J}\wedge \dd \a
\end{align}
in the NS-NS sector. For later convenience, we write $e^\phi\equiv g_s^\a$, $e^{\phi_0}\equiv g_s$ and 
$\gamma(\a)\equiv \frac{(5+\cos 2\a)^{3/4}}{(3+\cos 2\a)}$. With this notation, equation (\ref{dilaton}) becomes 
\be 
g_s^\a= g_s \gamma(\a)\; .
\ee
We note that the string coupling $g_s^\a$ is maximum at $\a=\pi/2$ and minimum at $\a=0,\pi$ (its derivative with respect to $\a$ is positive 
on the interval $[0,\pi/2]$ and negative on the interval $[\pi/2,\pi]$).

The RR fields are \footnote{Observe we correct a typo in the $4$-form potential of the original reference \cite{Guarino:2015jca}. We  thank Georgios Itsios for pointing out this mistake.}
\begin{align}
\widetilde{F}_0&=3^{-1/2}L^{-1}e^{-5\phi_0/4}\equiv m\\
\widetilde{F}_2&=-\sqrt{6} L e^{-3\phi_0/4}\left\{ \frac{4\sin^2\a \cos \a}{(3+\cos 2\a)(5+\cos 2\a)}{\cal J}
+\frac{3(3-\cos 2\a)}{(5+\cos 2\a)^2}\sin \a\; \dd \a \wedge \eta \right\}\\
&=-\sqrt{6} L e^{-3\phi_0/4}\left\{ \frac{4\sin^2\a \cos \a}{(3+\cos 2\a)(5+\cos 2\a)}{\cal J}-\dd \left(\frac{3\cos \a}{5+\cos 2\a}\right) \wedge \eta \right\}\\
\widetilde{F}_4&=L^3 e^{-\phi_0/4} \left\{ 6vol(AdS_4)-12\sqrt{3}\frac{(7+3\cos2\a)}{(3+\cos2\a)^2}\sin^4\a\; vol(\mathbb{CP}^2) \right.\nonumber\\
&\left.+18\sqrt{3}\frac{(9+\cos 2\a)\sin^3\a \cos \a}{(3+\cos 2\a)(5+\cos 2 \a)}{\cal J}\wedge \dd \a \wedge \eta\right\}
\end{align}
in the R-R sector. Finally, the constants are
\be 
L^2\equiv 2^{-5/8}3^{-1}g^{-25/12}m^{1/12}\; \quad \text{and} \quad e^{\phi_0}\equiv 2^{1/4}g^{5/6}m^{-5/6}\; .
\ee
\ese

Using the definitions of \cite{Bergshoeff:2001pv}, where $H_3=\dd B$ , $\widetilde{F}_2=\dd C_1+\widetilde{F}_0 B$ and $\widetilde{F}_4=\dd C_3+C_1\wedge H_3+\frac{\widetilde{F}_0 }{2} B \wedge B$, the potentials that generate these $p$-forms are given by \cite{Varela:2015uca}
\bse
\begin{align}
B & =-\frac{6L^2 e^{\phi_0/2}\sqrt{2}\sin^2\a \cos \a}{(3+\cos2\a)}{\cal J} - \frac{3 L^2 e^{\phi_0/2}}{\sqrt{2}}\sin\a \dd \a\wedge \eta \\
C_1 &=\frac{L e^{-3\phi_0/4}\sqrt{6}\sin^2\a \cos\a}{(5+\cos2\a)}\eta\equiv 2 a_0(\a)\eta\\
C_3 &=  L^{3} e^{-\phi_0/4} \left(6\Gamma + \frac{6\sqrt{3}\sin^4\a}{(3+\cos 2 \a)}{\cal J}\wedge \eta\right)\; ,
\end{align}
where $\dd \Gamma=vol(AdS_4)$, then $\Gamma=-\frac{1}{3}r^3 \sin \vartheta \dd t \wedge \dd \vartheta \wedge \dd \varphi $.
\ese
Furthermore, the field strengths satisfy the Bianchi identity
\begin{align}
\dd H_3=& 0 \\
\dd F_2= & \widetilde{F}_0 H_3\\
\dd F_4= & H_3 \wedge  \widetilde{F}_2\; .
\end{align}

More generally, we define the gauge-invariant field strength $\widetilde{F}_p$ as  
\be 
\widetilde{F}_p= \left( \dd C - H_3 \wedge C + F_0 e^B \right)_p\; ,
\ee 
and we see that the $p$-forms in the RR-sector are not closed, but satisfy the modified Bianchi identity $\dd \widetilde{F}_k= H_3 \wedge \widetilde{F}_{k-2}$, that is, the $p$-forms are closed under the derivative $\dd_H= \dd - H_3\wedge$.


\subsection{Quantization of charges}

We can consider the most appropriate flux in a supergravity background by the so called, \emph{Page charges} \cite{Marolf:2000cb}, 
defined as follows. Consider the $d$-closed $p$-form defined as 
\be 
\widehat{F}_p:= \left[ e^{-B}\wedge \left( \widetilde{F}_0 + \widetilde{F}_2 + \widetilde{F}_4 + \widetilde{F}_6 + \widetilde{F}_8 \right) \right]_p\; .
\ee 

Then the Page charges $N_k \in \mathbb{Z}$ are defined by
\be 
n_{p}:=\frac{1}{(2\pi \ell_s)^{p-1}}\int_{\Sigma^p} \widehat{F}_p\; ,
\ee
where $\Sigma^p$ is a closed space-like $p$-cycle and $l_s=\sqrt{\a'}$. In particular, 
\be
k\equiv n_0= 2\pi \ell_s F_0=2\pi \ell_s m.
\ee
is an integer that has a simple interpretation in the field theory as a CS level. 
In the solution considered in the last section, we have, on a 6-cycle (the whole compact space)
\be 
N\equiv n_6 =\frac{1}{(2\pi \ell_s)^5}\int_{{\cal S} S^5} \left( \widetilde{F}_{(6)} - B\wedge \dd C_{(3)}
-\frac{\widetilde{F}_{(0)}}{6}B\wedge B\wedge B \right)\;,
\ee
where $\widetilde{F}_{(6)}=-e^{\phi/2}\ast \widetilde{F}_4$, and the Hodge dual is taken with relation to the Einstein frame metric (\ref{metric}), as the authors of \cite{Guarino:2015jca} have shown. 
This integer will correspond to the rank of the gauge group in the field theory. 
In fact, these are the quantized charges that we have associated to the massive supergravity above, and the quantization of the fluxes implies that 
\bea
L&=& \frac{\pi^{3/8} \ell_s}{2^{7/48}3^{7/24}}(kN^5)^{1/24} ;\qquad
e^{\phi_0}=  \frac{2^{11/12}\pi^{1/2}}{3^{1/6}}\frac{1}{(k^5 N)^{1/6}} \Rightarrow \cr
L^2_{string}&=&\frac{2^{1/6} \pi}{3^{2/3}}\left(\frac{N}{k}\right)^{1/3}\ell_s^2\sqrt{5+\cos 2\a}\; .\label{quant}
\eea
as the authors \cite{Fluder:2015eoa, Varela:2015uca} have shown.

Note that 
\be
L_{string} e^{\phi_0}=\frac{2\pi \ell_s}{\sqrt{3}k}[5+\cos 2\a]^{1/4}\Rightarrow L_{string}g_s^\a=\frac{2\pi \ell_s}{k}\frac{5+\cos 2\a}{\sqrt{3}(3+\cos 2\a)}.
\label{quantrel}
\ee

In \cite{Marolf:2000cb}, the author showed that there are three distinct notions of charges associated to a given gauge field, 
namely, \emph{Brane charges, Maxwell charges} and \emph{Page charges}, and only the latter is quantized, as we have seen above. 

Consider, for example, the $4$-form field strength $\widetilde{F}_4=\dd C_3+ C_1 \wedge H_3 + \frac{F_0}{2}B\wedge B$, that satisfies 
the Bianchi identity $\dd \widetilde{F}_4= H_{3}\wedge\widetilde{F}_2 $. The first type of charge is given by the integral over the space 
${\cal M}^5$ such that $\partial {\cal M}^5= \Sigma^4$ of the current
\bse
\be 
\star {\cal J}_{D4}=\dd \widetilde{F}_4-H_3\wedge \widetilde{F}_2 \; ,
\ee
that is
\be 
Q_{D4}  = \frac{1}{8\pi^3\a'^{3/2} }\int_{{\cal M}^5} \star {\cal J}_{D4}\; ,
\ee
\ese
and this charge is gauge invariant but it is not quantized.

Similarly, we can define charges\footnote{Defined as $ \left(2\pi \ell_s \right)^{7-p} Q_{Dp}= \int_{{\cal M}^{9-p}}\star {\cal J}_{Dp} $.} 
for the D$2$-branes and D$6$-branes, through the following currents
\bse
\begin{align}
\star {\cal J}_{D6} & = \dd \widetilde{F}_2- \widetilde{F}_0 H_3 \phantom{x x} \;  \Rightarrow\;  Q_{D6}  = \frac{1}{2\pi\a'^{1/2} }\int_{{\cal M}^3} \star {\cal J}_{D6}\\
\star {\cal J}_{D2} & = \dd \widetilde{F}_6- H_3 \wedge \widetilde{F}_4 \;  \Rightarrow\;  Q_{D2}  = \frac{1}{32\pi^5\a'^{5/2} }\int_{{\cal M}^7} \star {\cal J}_{D2}\; ,
\end{align}
\ese
where $\widetilde{F}_6= - \star \widetilde{F}_4 $ .

We can define another important charge by regarding the right hand side of the equation
\bse
\be 
\dd \widetilde{F}= H_3\wedge \widetilde{F}
\ee
as a source for the field strength $\widetilde{F}$, so that
\be 
\dd \widetilde{F}_{8-p}=\star {\cal J}_{Dp}^{Maxwell}\; .
\ee
Integrating this equation and using the Stokes theorem, with the appropriate normalization, we find
\be 
Q_{Dp}^{Maxwell}=\frac{1}{(2\pi \ell_s)^{7-p}}\int_{\Sigma^{8-p}}\widetilde{F}_{8-p}\; .
\ee
\ese

In the Guarino-Jafferis-Varela (GJV) solution, we can easily see that 
\be 
n_6= Q_{D2}^{Maxwell} -\frac{1}{(2\pi \ell_s)^5}\int_{{\cal S} S^5} \left(B\wedge \dd C_{(3)}+\frac{\widetilde{F}_{(0)}}{6}B\wedge B\wedge B \right)
\ee
and using the previous results, we find
\be 
Q_{D2}^{Maxwell} = \frac{81\sqrt{6} \pi^3 e^{\phi_0/4}L^5 (4-\pi)}{(2\pi \ell_s)^5} .
\ee

Under this perspective, there is one more relevant cycle we may consider in the GJV solution, namely, the $4$-cycle parametrized by the 
coordinates $\Sigma^4=(\lambda, \theta, \phi, \sigma)$. We can easily see that this cycle is the $\mathbb{CP}^2$ metric multiplied by the 
factor $e^{\phi/2} \Delta \Xi$ that vanishes at the points $\a=0,\pi$, and since $\left.\widetilde{F}_4\right|_{\Sigma^4}$ also vanishes at 
these points, we conclude that the Page charge $n_4$ is zero in this cycle. On the other hand, the Maxwell charge is not trivial, that is
\be
\begin{split} 
Q_{D4}^{Maxwell} & =\frac{1}{(2\pi \ell_s)^3}\int_{\Sigma^4}\widetilde{F}_4\\
& = \frac{3\sqrt{3}}{8 \pi} \frac{L^3 e^{-\phi_0/4}}{\ell^3_s} \frac{(7+ 3\cos 2 \a)}{(3+ \cos 2\a)^2}\sin^4 \a
\; .
\end{split}
\ee

Furthermore, the Page charges are well defined up to a large gauge transformations, that shifts the Kalb-Ramond flux 
\be 
b=\frac{1}{2\pi \ell_s}\int_{\Sigma^2} B
\ee 
by an integer. In fact, since $\dd \widetilde{F}_2=\widetilde{F}_0 H_3$, we can write
\be 
B=\frac{\widetilde{F}_2}{\widetilde{F}_0}+ \mathcal{B}\; ,
\ee
where $\mathcal{B}=-\dd C_1/\widetilde{F}_0$ and we can easily verify that $\widetilde{F}_0 {\cal B}= - \widehat{F}_2$, so in principle 
it should be quantized, but once we consider a $2$-cycle $\Sigma^2$ with legs in the coordinate $\a$, the integral $\int_{\Sigma^2}\dd C_1$, 
vanishes trivially, as well as the Maxwell charge associated to this cycle \cite{Bergman:2010xd, Apruzzi:2013yva, Apruzzi:2015zna, Rota:2015aoa}. 
In particular, we can write the Kalb-Ramond flux as
\be 
b= \frac{2 \pi \ell_s}{n_0}\left( n_2 + Q_{D6}^{Maxwell} \right)\; ,
\ee
and if the Maxwell charge $Q_{D6}^{Maxwell}$ vanishes, the flux $b$ goes with $n_2/n_0$.

\subsection{Field theory dual}

The field theory dual to this gravitational solution was proposed to be the maximal ${\cal N}=8$ SYM theory in 2+1 dimensions {\em with a single 
gauge group $SU(N)$}, the $N$ corresponding to the integer in the quantized flux on $S^6$ (the Page charge $n_6$ of the previous subsection), 
deformed by Chern-Simons terms with level $k$ related to the mass parameter $m$ by the relation 
\be
m=\frac{k}{2\pi \ell_s}.
\ee
As we saw, $k$ is the same as the Page charge $n_0$ of the previous subsection. 

The field theory has 7 adjoint scalars (corresponding to the 7 transverse coordinates to the 2+1 dimensional worldvolume in 10 dimensional string theory),
and 8 fermions transforming under the R-symmetry group $SO(7)$. At low energy, an adjoint scalar dual to a photon completes the fundamental 
representation of the full $SO(8)$ R-symmetry group. 

The addition of the Chern-Simons term
\be
\frac{k}{4\pi}\Tr\left[A\wedge F+\frac{2}{3} A\wedge A\wedge A\right]\;,
\ee
together with additional couplings, can preserve ${\cal N}=2$ supersymmetry, like the gravitational solution. 
The SYM theory in 3 dimensions has a dimensionful coupling, but the theory flows in the IR  to a fixed point with  
conformal symmetry, corresponding to the warped $AdS_4$ factor in the gravity dual. 
In this fixed point, the effective 't Hooft coupling is 
\be
\lambda=\frac{N}{k}\;,
\ee
which must be $\gg 1$ for the supergravity approximation to hold in the gravity dual. Indeed, the curvature in string units is large only if $N/k\gg 1$, 
as can be see from (\ref{quant}).

In ${\cal N}=2$ notation, we have an adjoint vector multiplet (that contains a real scalar and a complex fermion) 
and 3 chiral multiplets with complex scalars $Z,W,T$, 
like the dimensional reduction of the 3+1 dimensional ${\cal N}=4$ SYM, with the same superpotential,
\be
W=g\Tr(Z[W,T])\;,
\ee
but where $g$ has dimension 1/2, so in the IR we must have $[\phi_i,\phi_j]=0$. There is also a conformal potential term for the scalars
\be
V_c=\frac{4\pi^2}{k^2}\Tr\left([[\bar\phi^i,\phi^i],\bar\phi ^k][[\bar\phi^j,\phi^j],\phi^k]\right).
\ee

The theory has R-symmetry $U(1)\times SU(3)$, with $SU(3)$ rotating the 3 complex scalars in the chiral multiplets. 

The symmetries correspond to the symmetries of the gravity solution. The would-be $SO(7)$ symmetry of the $S^6$ is also broken by the 
squashing/fibration to $SU(3)$, acting on $\mathbb{CP}^2$, times $U(1)$, translating $\psi$, or rather $\eta$.


\section{Particle-like branes and their field theory duals}

In this section we perform a qualitative analysis of branes wrapped on cycles, that look like particles from the point of view of the field theory, and their dual intepretation.
We will study static D-branes and moving D-branes. 

The D-brane action is inversely proportional to the string coupling. If $g_s$ would get large, we would have light D-brane states.
But in fact, in \cite{Aharony:2010af}, the authors argued that massive type IIA theory cannot be strongly coupled. In the examples considered in their 
work, the authors showed that the string coupling $g_s$ has two regimes. In the first regime, $g_s$ increases with $N$ as usual in the massless
 theories, but then it reaches a second regime where it decreases for large $N$. 

In the GJV solution, we can easily prove that there is no such phase transition,  since the coupling $g_s^\a$ remains bounded, 
\be 
g_s^\a=e^{\phi}\sim \frac{1}{k^{5/6} N^{1/6}}< \frac{\ell_s}{L_{string}}\; , \quad \forall \ N,k\;, \label{coupling}
\ee
where from (\ref{quant}), $L_{string}\sim N^{1/6} k^{-1/6}\ell_s$ (see also (\ref{quantrel})), and moreover as we said, must be taken to be $\gg \ell_s$ 
for validity of the supergravity approximation, by imposing $N\gg k$.

\subsection{Branes on cycles in the GJV geometry}\label{sec:brane}

In this subsection we consider static branes, that will correspond to solitonic operators in the dual field theory. 

We can write the action for a general D$p$-brane as the sum of the DBI term and a WZ term,
\be 
S_{Dp}=-T_{Dp}\int \dd^{p+1}\xi e^{-\phi}\sqrt{-\det(G+{\cal F})} - 
T_{Dp} \int \sum_{\tilde p} C_{\tilde p}\wedge e^{\cal F}\; ,
\ee
where ${\cal F}=B+ 2\pi \a' F$ and $T^{-1}_{Dp}=(2\pi)^p \ell_s^{p+1}$. It has been shown in \cite{Green:1996bh} that in the presence of 
Romans mass, we need to add the  WZ term for $F_0$,
\be 
\begin{split}
S_{m}^{Dp} & =  T_{Dp} F_0 \left. \int \sum_{r=0}(2\pi\a')^r\frac{\omega^0_{2r+1}}{(r+1)!}\right|_{p+1}\; \label{cs-mass}\\
& =  \frac{k T}{(2\pi)^{p} \a'^{p/2} } \left. \int \sum_{r=0}(2\pi\a')^r\frac{\omega^0_{2r+1}}{(r+1)!}\right|_{p+1}\;,
\end{split}
\ee
where $\omega_{2r+1}^0$ is a Chern-Simons form, such that $\dd \omega^0_{2r+1}=\tr (F)^{r+1} \Rightarrow \omega^0_{2r+1}= 
\tr \left(A\wedge F^r\right)$, and in the second expression we have used that  $F_0=n_0 T \sqrt{\a'}\equiv kT\sqrt{\a'}$, 
where $T=1/(2\pi \a')$ is the string tension and $F=dA$. 
We consider just an abelian ($U(1)$) field strength $F$. 

{\bf D0-branes}

We start with the analysis of D0-branes, which therefore need not be wrapped on cycles, since they are already particle-like. 
The D$0$ brane action develops a tadpole from the WZ term (\ref{cs-mass}), namely
\be 
S_{m}^{D0}= -T_{D0} F_0\int_{\mathbb{R}} \dd t A_t= -n_0 T \int\dd t A_t\; .
\ee
That means that we need to add $k$ fundamental strings ending on the D$0$-brane. 
In principle, there is also the leading term in the WZ part, $\int_\Sigma C_1$, but since $C_1$ has no component along $dt$, corresponding to 
$\Sigma=\mathbb{R}_t$, for a {\em static} D0-brane as we consider here, this contribution vanishes. 

Moreover, from the DBI action
\be 
S_{DBI}=-T_{D0}\int_{\mathbb{R}} \dd t \ e^{-\phi} \sqrt{|G_{tt}|} \; ,
\ee
we easily see that the brane mass\footnote{Given a static D$p$-brane wrapping a $p$-cycle $\Sigma$ with a vanishing WZ term, 
its mass is given by the DBI term, $M_{Dp}=T_{Dp}\int_{\Sigma} 
g_s^{-1} \sqrt{ \det_{\Sigma} (G+{\cal F}) }$, where $\det_{\Sigma}$ means the determinant along the $p$-cycle $\Sigma$.} 
is 
\be
M_{D0}(\a)=\frac{T_{D0}}{g_s^\a}\sim T_{D0}(N k^5)^{1/6}\gamma(\a)^{-1}.
\ee
Since as we saw, the string coupling $g_s^\a$ is maximal at $\a=\pi/2$, where it takes the value $e^{\phi_0}\sqrt{2}$, 
the D0-brane will stabilize at this value of $\a$, that minimizes its energy, giving
\be
M_{D0}=T_{D0}k\left(\frac{N}{k}\right)^{1/6}\frac{3^{1/6}}{2^{17/12}\pi^{1/2}\sqrt{\pi}} .\label{dualbmass}
\ee

In conclusion, this D0-brane needs $k$ fundamental strings to end on it. 

{\bf D6-branes}

The D$6$-brane wrapping the whole internal space $\mathcal{S} S^5$ develops a tadpole due to the WZ coupling
\be 
S_{D6}^{WZ}=-T_{D6} \int_{\mathbb{R}\times\mathbb{{\cal S} S^5}} \widehat{F}_6\wedge A=-N T \int_{\mathbb{R}} \dd t  A_t\;,
\ee
thus we need $N$ fundamental strings ending on this brane. The leading WZ coupling for the D6-brane would come from $C_7$
integrated over the compact space, where 
$\tilde F_8=dC_7-H_3\wedge C_5+...=-e^{\phi/2}\tilde F_2$, but since $\tilde F_2$ has only components in the compact dimensions, 
$\tilde F_8$ is nonzero only when it has at least two non-compact directions. The same is true for the terms $H_3\wedge C_5+...$, 
which means that the integral of $C_7$ over the compact space and time is zero.

The mass of the D$6$-brane is then given by
\be 
M_{D6} =T_{D6}\int \dd^6 \xi e^{-\phi}\sqrt{ G_6}= 
\sqrt{\frac{3}{2}}\pi^3 T_{D6} \int_0^\pi \dd \a \frac{\sqrt{\Omega} \Xi^2 L^6_{string}}{e^{\phi}} \sim (T_{D6}\a'^3)\frac{N^{7/6}}{k^{1/6}} \; ,\label{d6-mass}
\ee
where $G_6$ is the determinant of the internal manifold in (\ref{metric}). More precisely, one obtains (also taking into account that 
$T_{D6}(2\pi \sqrt{\a'})^6=T_{D0}$)
\be
M_{D6}=T_{D0}N\left(\frac{N}{k}\right)^{1/6}\frac{3^{5/3}}{2^{59/12}\sqrt{\pi}}\int_0^\pi d\a \sin^5\a \frac{(5+\cos 2\a)^{1/4}}
{3+\cos 2\a}.\label{baryonmass}
\ee

An observation about the calculation in (\ref{d6-mass})
is that the warp factor $L_{string}$ is $\a$-dependent, and since we still have one factor $L_{string}$ from the component 
$G_{tt}$ in the metric, the Dirac-Born-Infeld is not the product of (\ref{d6-mass}) and the time integration of $\sqrt{G_{tt}}$. 
Moreover, for the sake of simplicity, we have ignored the $B$-field.


{\bf Bound state of D0-branes and D6-branes}

Considering a bound state consisting on $p$ D$0$-branes and $q$ D$6$-branes, the tadpole is
\be 
\left( p k + q N \right)\int A\; ,
\ee 
and this term vanishes when $p=-\frac{N}{k}q$. The leading term in the WZ piece is zero, as we just saw, so the mass of this system is
\be 
M_{D0-D6}=\sqrt{p^2 M_{D0}^2+q^2 M_{D6}^2}.\label{boundstate}
\ee
In the case of $p\sim q$ and $N\gg k$ needed for the validity of the supergravity approximation, we can write
\be
M_{D0-D6}\simeq q M_{D6}\sim q \frac{L_{string}^6}{g_s}\propto qN \left(\frac{N}{k}\right)^{1/6}\; .
\ee
But in general, the formula is symmetric under the interchange of $M_{D0}$ with $M_{D6}$ and $p$ with $q$. 

{\bf D2-branes and D4-branes}

In addition, we see that D$2$-branes and D$4$-branes are tadpole free, since if we consider that D2-branes on $\mathbb{CP}^1\subset\mathbb{CP}^2$
and D4-branes on $\mathbb{CP}^2$, the $\a$-dependent  Wess-Zumino terms 
\be 
\int_{\mathbb{CP}^1\times \mathbb{R}_t} \dd C_1 \wedge A\;\quad \text{and}  \quad \int_{\mathbb{CP}^2\times \mathbb{R}_t} \dd C_3 \wedge A
\ee
will be zero at the minimum. Indeed, 
\bea
\int_{\mathbb{CP}^1} dC_1&=&a_0(\a)\int_{\mathbb{CP}^1}2{\cal J}=-2\pi a_0(\a)\propto \sin^2\a\cos \a\cr
\int_{\mathbb{CP}^2}dC_3&=&12\sqrt{3}L^3e^{-\phi_0/4}\frac{\sin^4\a}{3+\cos 2\a}\int_{\mathbb{CP}^2}{\cal J}\wedge {\cal J}\propto 
\sin^4\a\;,
\eea
so are both minimized by $\a=0$, where they take the value zero.
In the absence of magnetic flux, these are the only tadpoles that could appear. 

 The leading term in the WZ piece for the D2-brane, coming from $\int C_3$, is trivially zero, but can be made nonzero by considering 
 a moving trajectory, $\eta=\eta(t)$ (so that the rest is $\int_{\mathbb{CP}^1}{\cal J}=-\pi$). The leading term in the WZ piece for the D4-brane, 
 coming from $\int C_5$, is also trivially zero, but can be made nonzero by a moving trajectory, again $\eta=\eta(t)$ (or rather, $\psi=\psi(t)$). 
We will analyze these later.

{\bf Branes with worldvolume magnetic flux}

Instead, we can consider a worldvolume magnetic flux given by 
$F={\cal N\;  J}$ in $S^2\equiv \mathbb{CP}^1\in \mathbb{CP}^2$, where ${\cal J}$ is the K\"ahler form of $\mathbb{CP}^2$, at fixed $\lambda$, 
and ${\cal N}\in 2 \mathbb{Z}$, see \cite{Janssen:2006sc, Gutierrez:2010bb, Lozano:2011dd}. For now, we ignore the subtle issue of the 
Freed-Witten anomaly \cite{Freed:1999vc, Aharony:2009fc}.

 The {\bf D$2$-brane} wrapping $\mathbb{CP}^1$, for any $\a$ (including $\a=0$), develops the coupling
(from the $\int F_0\wedge F\wedge A$ WZ term)
\be 
-\frac{k {\cal N}}{2} T\int_{\mathbb{R}} \dd t A_t\; ,
\ee
where we have used the fact that $\int {\cal J}=-\pi$ in $\mathbb{CP}^1$,
so that now $k{\cal N}/2$ fundamental strings are required to cancel this tadpole.

The magnetic flux also induces a dissolved D0-brane charge of ${\cal N}$ from the usual WZ coupling
\be
T_{D2}\int_{\mathbb{CP}^1}F\wedge \int _{\mathbb{R}_t}C_1=-{\cal N}T\int _{\mathbb{R}_t}C_1.
\ee

Moreover, the DBI term in the mass of the D$2$-brane wrapping 
the squashed space $\widetilde{\mathbb{CP}}^1$ is 
\bse
\be 
\begin{split}
M_{D2}(\a)&= \frac{T_{D2}}{g_s^\a}\int_{\widetilde{\mathbb{CP}}^1} \sqrt{\det (G_{\a\b} + 2\pi \a' F_{\a\b})}\\
&= \frac{\pi T_{D2}}{2g_s^\a}\int_0^\pi \dd \theta \sqrt{\left[(\Xi^2-\Xi\Omega)L_{string}^4 + \left(\pi \a' {\cal N}\right)^2\right] \sin^2\theta + 
L_{string}^4\Xi\Omega }\; ,
\end{split}
\ee
which is an elliptic integral. For the sake of simplicity, we ignore $\Omega$ , then 
\be 
\begin{split}
M_{D2}(\a)&= \frac{T_{D2}}{g_s^\a}\int_{\mathbb{CP}^1} \sqrt{\det (G_{\a\b} + 2\pi \a' F_{\a\b})}\\
&= \frac{\pi T_{D2}}{g_s^\a}\left(\Xi^2L_{string}^4 + \left(\pi \a' {\cal N}\right)^2 \right)^{1/2}\; ,
\end{split}
\ee
\ese
and we see that the mass of the brane has the usual term related to the brane wrapping an internal cycle, and a shift related to the magnetic charge ${\cal N}$.
The first (flux-independent) term is proportional to 
\be
\sin^4\a(5+\cos 2 \a)^{1/4}\;,
\ee
so at ${\cal N}=0$, the whole mass (WZ term plus DBI) is minimized by $\a=0$, at value $M=0$. The second term however, 
is proportional to
\be
\frac{3+\cos 2\a}{(5+\cos 2\a)^{3/4}}\;,
\ee
which is minimal at $\a=\pi/2$ and maximal at $\a=0$. Together, at ${\cal N}\neq 0$, the mass of the D2-brane (DBI plus vanishing WZ term) is 
minimized at $\a\neq 0$, for a nonzero mass.

The {\bf D$4$-brane} wrapping the $\mathbb{CP}^2$ cycle at nonzero magnetic flux has a tadpole coming from
\be 
-\frac{k {\cal N}^2}{8} T \int_{\mathbb{R}}\dd t A_t\; ,
\ee
and we need $k {\cal N}^2/8$ fundamental strings to cancel this tadpole. Observe that now we cannot guarantee that $k {\cal N}^2/8$ is an integer, 
since ${\cal N}^2\in 4 \mathbb{Z}$. Similar to what occurs in the analysis \cite{Gutierrez:2010bb}, this suggests that a fractional number of 
fundamental strings should be added to the D$4$-brane worldvolume to cancel this tadpole. Moreover, we have a shift in the value of the 
D$4$-brane mass due to the magnetic field $F={\cal N}{\cal J}$, giving
\be 
M_{D4}(\a)=\frac{\pi^2 T_{D4}}{2 g_s^\a}\left(\Xi^2 L_{string}^4 + (\pi\a' {\cal N})^2 \right)\; ,
\ee
where we have deliberately ignored the squashed term $\Omega$ in the metric \footnote{In fact, the complete expression has the square root 
of the following determinant
\begin{align*}
\det(G+2\pi \a' F) &= 
\frac{1}{128} \sin ^2\theta  \sin ^6\lambda \left(L_{string}^4 \Xi^2+\pi^2 \alpha'^2 {\cal N}^2\right)\times\\
& \times \left[\cos 2 \lambda  \left(L_{string}^4 \Xi  (\Xi -\Omega )+\pi^2 \alpha'^2 {\cal N}^2\right)+L_{string}^4 \Xi  (\Xi +\Omega )
+\pi^2 \alpha'^2 {\cal N}^2\right]\; .
\end{align*}
}.
Exactly the same comments as for the D2-brane case apply for the minimization with respect to $\a$.

The D4-brane on $\mathbb{CP}^2$ also has an induced dissolved D2-brane charge coming from the WZ term
\be
S=2\pi T_{D4}\int_{\mathbb{R}_t\times \mathbb{CP}^2}C_3\wedge F=\frac{\cal N}{2}T_{D2}\int C_3
\ee
and a dissolved D0-brane charge coming from the WZ term
\be
S=\frac{1}{2}(2\pi)^2T_{D4}\int_{\mathbb{R}_t\times \mathbb{CP}^2}=\frac{{\cal N}^2}{8}D_{D0}\int_{\mathbb{R}_t}C_1.
\ee

Finally, the {\bf D$6$-brane} does not have any tadpole related to the Romans mass $F_0$, since ${\cal J\wedge J\wedge J}=0$. The reader 
should compare this result with \cite{Gutierrez:2010bb}, where the Romans mass contributes to the tadpole in a deformation of the ABJM 
theory \cite{Gaiotto:2009mv}. On the other hand, the DBI contribution to the  mass of the D$6$-brane is given by
\bse
\be 
M_{D6} =T_{D6}\int \dd^6 \xi e^{-\phi}L^2_{string}\frac{ \sqrt{ \det(G_{{{\cal S} S^5}})}}{\Xi^2} \left(L^4_{string} \Xi^2 + (\pi \a' {\cal N})^2 \right) \; ,
\ee
where $\det(G_{{{\cal S} S^5}})$ is the determinant (\ref{det}), so
\be 
\begin{split}
M_{D6} & = \frac{\sqrt{3}}{8\sqrt{2}} T_{D6} \int \dd^6 \xi e^{-\phi}\sqrt{\Omega} L^2_{string} \cos \lambda \sin^3\lambda \sin \theta
 \left(L^4_{string} \Xi^2 + (\pi \a' {\cal N})^2 \right)\\
&=\sqrt{\frac{3}{2}}\pi^3 T_{D6} \int_0^\pi \dd \a \frac{\sqrt{\Omega} L^2_{string}}{g_s^\a} \left(L^4_{string} \Xi^2 + (\pi \a' {\cal N})^2 \right) \; ,
\end{split} 
\ee
\ese
and we see that the magnetic flux shifts the mass of the D$6$-brane wrapping the internal manifold. Moreover, now we have an integration 
over $\a$, so we don't need to minimize with respect to it. The contribution of the WZ term 
$\int C_7$, where $\tilde F_8=-e^{\phi/2}\tilde F_2$, is zero, since $C_7$ doesn't have a term that covers the whole compact space (and time). 
Therefore the mass is given entirely by the DBI term.

In this case again there are induced dissolved D-brane charges, D2-brane charge coming from 
\be
\frac{1}{2}T_{D6}\int C_3\wedge F\wedge F\;,
\ee
but no D0-brane charges, since 
\be
T_{D6}\int C_1\wedge F\wedge F\wedge F=0
\ee
gives zero (${\cal J}\wedge{\cal J}\wedge {\cal J}=0$).


\subsection{Field theory dual: baryon operators}

We would like to understand what these particle-like branes with strings ending on them mean from the point of view of the dual field theory. 

We start by summarizing the results for the various branes we have found, in table (\ref{table}).

\begin{table}[ht]
\begin{tabular}{|c|c|c|}
\hline 
Dp-branes & Mass $M_{Dp}$ & $q=\#$ Strings \\ 
\hline
\hline 
D$0$-branes & $T_{D0}/g_s^\a$ & $k$ \\ 
\hline 
D$2$-branes & $\pi T_{D2} \left(\Xi^2 L^4_{string} + (\pi\a' {\cal N})^2 \right)^{1/2}/g_s^\a$ & $k{\cal N}/2$\\
\hline
D$4$-branes & $\pi^2 T_{D4} \left(\Xi^2 L^4_{string} + (\pi\a' {\cal N})^2 \right)/2 g_s^\a$ & $k{\cal N}^2 /8$\\
\hline
D$6$-branes & $\sqrt{3/2}\pi^3 T_{D6}/g_s \int \dd \a \frac{\sqrt{\Omega} L^2_{string}}{\gamma(\a)} \left(L^4_{string} \Xi^2 + (\pi \a' {\cal N})^2 \right) $ & $N$\\
\hline
\end{tabular} 
\caption{Branes configurations.}\label{table}
\end{table}

The static solitonic branes with strings attached to them correspond to baryon vertex operators (when introducing external quarks in the theory), 
as understood first by Witten \cite{Witten:1998xy} in the context of 
the D5-brane wrapping $S^5$ in $AdS_5\times S^5$ for ${\cal N}=4$ SYM. More precisely, in that case for the gauge group $SU(N)$, one could 
construct the baryon 
\be
B^{I_1...I_N}=\epsilon_{i_1...i_N} q^{i_1I_1}...q^{i_NI_N}\;,
\ee
and the external (heavy) quarks $q^{i_kI_k}$ correspond to long strings coming from the boundary and reaching the brane. The mass of the brane 
(\ref{baryonmass}) corresponds to the mass of the baryon with the quark masses subtracted, i.e. the baryon vertex operator.

The field theory dual to the GJV solution is also a $SU(N)$ SYM theory, though now with a CS term at level $k$. As such, we again expect the 
presence of the same baryon vertex constructed using the $\epsilon_{i_1...i_N}$ tensor. We indeed find this from the D6-brane wrapped over the 
compact space, which indeed corresponds in the T-dual type IIB theory to the D5-brane wrapped over the compact space. 

The D2-brane and D4-brane baryons are harder to understand, but they appear only in the presence of magnetic flux, so they represent some 
interesting dynamical objects, that should be studied further. 

Also the D0-branes, giving a baryon made up of  $k$ external quarks, are harder to understand from the point of view of the $SU(N)$ 
gauge theory. But it is known that many Chern-Simons gauge theories have level-rank duality, as reviewed, derived and extended in the 
recent paper \cite{Hsin:2016blu}. In certain cases, an $SU(N)_k$ theory is dual to an $SU(k)_N$ theory \cite{Naculich:1990pa,Mlawer:1990uv},
or even in the case of $U(N)_k$ we can have a duality to $U(k)_N$ \cite{Naculich:2007nc}. In \cite{Hsin:2016blu}, a duality valid in general was proposed
between a $SU(N)$ theory at level $k$ and an $U(k)$ theory at level $-N$. 

In our case, the theory contains other fields besides the gauge fields, and also a SYM term, not just the CS term (the SYM term however
is subdominant to the CS term at low energies), so the physics is not very clear, but we would  expect from the level-rank  dual formulation to have
at the very least a baryon constructed with the $\epsilon_{i_1...i_k}$ tensor. Moreover, the mass of this baryon vertex must match (\ref{dualbmass}). 
An interesting observation is that both (\ref{dualbmass}) and (\ref{baryonmass}) are proportional to $T_{D0}(N/k)^{1/6}=T_{D0}\lambda^{1/6}$, whereas 
the D0-brane mass is proportional to $k$ and the D6-brane mass with $N$. This suggests that level-rank duality certainly plays a role, in exchanging 
the $N$ and $k$ prefactors, whereas the coupling dependence stays the same $\lambda^{1/6}$, and is not inverted as naive replacing of $N$ with 
$k$ would suggest.

Finally, we have also bound states of the D0-branes and D6-branes, giving new bound states of the two types of baryon vertices. Again from the 
point of view of the field theory, we have evidence of level-rank duality, when we exchange $N$ with $k$, $p$ with $q$, while leaving 
$\lambda=N/k$ invariant, as seen in (\ref{boundstate}). The case of $p=-\frac{N}{k}q$ (the minus sign signifies considering anti-baryons) is 
very interesting: in this case there is no tadpole, so it would seem like we don't need external quarks, but rather have a solitonic object 
composed of only glue, and yet with the symmetry properties of a combination of baryons. Unlike the case of a usual $N$-baryon/$N$-antibaryon 
combination, where the result would be annihilation, in this case the bound state would seem to be stable, which is quite unusual. 

We see that we have strong evidence for level-rank duality in the field theory and, since such a relation is expected for our theory, this also 
is strong evidence for the correct identification of the field theory dual to the gravitational GJV solution. 

Another observation is that 
it is expected that the external magnetic flux breaks the level-rank duality, and indeed we see that the D2 and D4 baryonic vertices are not 
level-rank dual to each other (or to themselves).


\subsection{Brane stability and operator bounds}

Now we try to proceed as in \cite{Janssen:2006sc, Gutierrez:2010bb} and analyze the dynamical stability of the brane configurations of subsection
(\ref{sec:brane}). This will give bounds on the magnetic fluxes, that can be reinterpreted as bounds on the dissolved charges (since those depend 
only on ${\cal N}$, as we saw).

We want to consider the system of a D$p$-brane wrapping the $p$-cycle and $q$ fundamental strings necessary to cancel the tadpole, that is
\be 
S=S_{Dp} + S_{qFS}\; ,
\ee 
where we consider the string at fixed $\a$, defined by the minimum in the mass of the D$p$-brane, $\a_{min}$.
\footnote{In the D$6$-brane case, the extra $L_{string}$ must be included in the integral  over$\a$ defined in $M_{D6}$.}
\bse
\begin{align}
S_{Dp} & =-M_{Dp} \int \dd t\   L_{string} u \equiv -M_{Dp}^L \int \dd t\ u \\
S_{qFS} & = - q T L_{string}^2|_{\a,min.} \int \dd x \dd t \sqrt{u^4 + u'^2}
\end{align}
\ese
and we have considered a string configuration with $x=-\s, t=\tau$ and $u=u(\s)$. There are two contributions to the equations of motion of this 
system, one from the bulk and another from the boundary  \cite{Brandhuber:1998xy, Maldacena:1998im, Sonnenschein:1999if}. These equations are,
respectively,
\bse
\begin{align}
\d_\s\frac{\d\cal{L}}{\d x'}=0\ \Rightarrow & \ \frac{u^4}{\sqrt{u^4 + u'^2}} =c_0 \\
\text{Boundary} \ \Rightarrow &\ \frac{u'_0}{\sqrt{u^4_0x_0'^2 + u'^2_0}}  = \frac{M_{Dp}^L}{L^2_{string,\a,min.} q T}\; .
\end{align}
\ese
We define
\be 
\sqrt{1-\b^2}\equiv \frac{M_{Dp}^L}{L^2_{string,\a,min,} q T}\quad \Rightarrow\quad \b=\sqrt{1 - \left( \frac{M_{Dp}^L}{L^2_{string,\a,min.} 
q T} \right)^2} \; ,\label{beta}
\ee
so that we write the equations of motion for the D$p$-brane q-strings as one single equation
\be 
\frac{u^4}{\sqrt{u^4 + u'^2}} = \b u_0^2\; .
\ee
Integrating this equation, we find the size of the brane configuration,
\be 
l_{wb}=\frac{1}{u_0}\int_1^\infty \dd y \frac{\b}{y^2\sqrt{y^4-\b^2}}\; ,\label{size}
\ee
where $y=u/u_0$. This equation is similar to the equations in \cite{Janssen:2006sc, Gutierrez:2010bb}, and the 
integral is solved in terms of hypergeometric functions. 

{\bf D0-brane}

Consider first the D$0$-brane. Reality of the function (\ref{beta}) implies
\be 
0<\frac{M_{D0}^L}{L_{string,\a,min.}qT}=\frac{2\pi \ell_s}{g_{s,min.}^\a k L_{string,\a,min.}}<1\; ,
\ee
but this is always satisfied, because from (\ref{coupling}) we have 
\be 
1> \frac{\ell_s}{g_s^\a L_{string,\a}}>\frac{2\pi\ell_s}{g_s^\a L_{string,\a}k}.
\ee

Consider next D2-branes and D4-branes {\em with worldvolume magnetic flux}.

{\bf D2-brane}

Neglect the squashed term $\Omega$ in the metric. For the D$2$-brane, with $q=k{\cal N}/2$, 
reality of $\b$ in (\ref{beta}) implies
\be
\frac{M_{D2}}{L_{string,\a,min.}T}<\frac{k{\cal N}}{2}\;,
\ee
which gives, since $\pi T_{D2}/T=1/(4\pi \a')$,
\be
\left(\frac{\Xi L_{string,\a,min.}}{2\ell_s g_{s,min.}^\a}\right)^2+\left(\frac{k{\cal N}}{2}\right)^2\left(\frac{\sqrt{3}(3+\cos 2\a_{min.})}{2(5+\cos 2\a_{min.})}\right)^2
<\left(\frac{k{\cal N}}{2}\right)^2\;,\label{D2}
\ee
which is a lower bound on $q=k{\cal N}/2$, or (at fixed $k$) on the dissolved charge ${\cal N}$.  Parametrically, we obtain 
\be
(k{\cal N})^2>{\cal O}[N^{2/3}k^{4/3}]\Rightarrow
{\cal N}>{\cal O}\left[\left(\frac{N}{k}\right)^{1/3}\right].
\ee 
In \cite{Janssen:2006sc, Gutierrez:2010bb}, the authors showed that the there are stable 
D$2$-brane configurations in the presence of magnetic flux ${\cal N}$ until some upper bound, interpreted as an upper bound on dissolved charges, 
related to the stringy exclusion principle. In the case (\ref{D2})  however, we have a lower bound,which means that stability of this configurations 
implies the presence of a minimum magnetic flux on the brane worldvolume. 

{\bf D4-brane}

A similar situation occurs for the D$4$-brane, with $q=k{\cal N}^2/8$. Stability of the configuration requires that $\b$ in (\ref{beta}) is real, which 
implies
\be
\frac{M_{D4}}{L_{string,\a,min.}T}<\frac{k{\cal N}^2}{8}.
\ee
This in turn gives
\be
\frac{1}{2}\left[\frac{\Xi^2L^3_{string,\a,min.}/\ell_s^3}{8\pi g_s^\a}+\frac{k{\cal N}^2}{8}\frac{\sqrt{3}(3+\cos 2\a_{min.})}{2(5+\cos 2\a_{min.})}\right]
<\frac{k{\cal N}^2}{8}\;,
\ee
which is again a lower bound on $q=k{\cal N}^2/8$, or (at fixed k) on the dissolved D0-brane charge ${\cal N}^2/8$. It is then also a lower bound on 
the magnetic flux on the worldvolume that makes the D4-brane stable, parametrically also
\be
k{\cal N}^2>{\cal O}[N^{2/3} k^{1/3}]\Rightarrow {\cal N}>{\cal O}\left[\left(\frac{N}{k}\right)^{1/3}\right].
\ee

{\bf D6-brane}

Finally, we consider the case of D6-branes, with $q=N$. In this case there is a small difference: the coordinate $\a$ is integrated over on the D6-brane 
worldvolume, but the strings need to be at given $\a$. Minimizing their action means that their $L_{string}^2$ factor must be minimum, i.e. for 
$\a=\pi/2$, when 
\be
L_{string}(\pi/2) g_s=\frac{2\pi \ell_s}{k}\sqrt{\frac{2}{3}}.
\ee
  
The stability bound is again the reality of $\b$, which gives
\be
\frac{M_{D6}^L}{L_{string}^2(\pi/2)T}=\frac{M_{D6}}{L_{string}(\pi/2) T}<N\;,\label{D6bd}
\ee
and is rewritten as 
\be
\frac{k}{(2\pi \ell_s)^6}\pi^3\int d\a\frac{\sqrt{\Omega}L^2_{string}}{\gamma(\a)}[L^4\Xi^2+(\pi \ell_s {\cal N})^2]<N.
\ee

This is now really an upper bound on the magnetic flux ${\cal N}$, or more precisely on the induced dissolved D2-brane charge $\sim {\cal N}^2$,
just like in \cite{Janssen:2006sc, Gutierrez:2010bb}. In order of magnitude, we obtain
\be
k\left[{\cal O}\left(\frac{N}{k}\right)+{\cal N}^2{\cal O}\left(\left(\frac{N}{k}\right)^{1/3}\right)\right]< N\Rightarrow
{\cal N}^2<{\cal O}\left[\left(\frac{N}{k}\right)^{2/3}\right].
\ee

In conclusion, we see that D0-branes are always stable, but the regions of stability of the D2-branes and D4-branes, and of the D6-branes 
are parametrically opposite: ${\cal N}>{\cal O}[(N/k)^{1/3}]$ and ${\cal N}<{\cal O}[(N/k)^{1/3}]$, respectively. There might be a window for all of 
them to be stable, depending on the exact numerical factors, which one would have to compute numerically.


\subsection{Reducing the number of fundamental strings}

Following \cite{Brandhuber:1998xy, Lozano:2011dd} we can consider a configuration with a reduced number $r$ of fundamental 
strings going to the boundary, the rest of the strings (up to the number required to cancel the tadpole) can go to the IR of the gravity solution. 
In holographic terms, this means that the baryons we consider have less quarks than the baryons of the previous subsections.

Using the conventions of \cite{Lozano:2011dd}, the configuration we consider consists of a D$p$-brane wrapped on a $p$-cycle and located at 
$u=u_0$ in the radial direction. In addition, we consider $l$ strings stretching from $u_0$ to the boundary of AdS4 (parametrized in polar 
coordinates) and $(q-l)$ straight strings
stretching from $u_0$ to $u=0$. The action is given by
\be 
S=S_{DBI}+ S_{lFS}+ S_{(q-l)FS}
\ee
where
\bse
\begin{align}
& S_{DBI}^{Dp}=- M_{Dp}^L\int \dd t u\\
& S_{l F1}  = -l T_{F1} L_{string}^2 \int \dd t \dd u \sqrt{1+ u^4 r'^2}\\
& S_{(q-l) F1}  = -(q-l) T_{F1} L_{string}^2 \int \dd t \dd u\; .
\end{align}
\ese
The equation of motion for the $l$ strings implies that
\be 
r'=\frac{\tilde{u}^2}{u^2\sqrt{u^4-\tilde{u}^4}}\; ,\label{r-eq}
\ee
where $0 \leq \tilde{u}\leq u_0 $ is the turning point of each string, and we see that at the turning point $r'\to \infty$. Moreover, as in 
\cite{Lozano:2011dd}, the boundary equation can be written as
\be\label{boundary-eq} 
\frac{1}{a}\sqrt{1-\tilde{\b}^2}+\frac{1}{a}(1-a)=\sqrt{1-\frac{\tilde{u}^4}{u_0^4}}\; \leq 1
\ee
where $a=l/q$ and
\be 
\sqrt{1-\tilde{\b}^2}=\frac{M_{Dp}^L}{L_{string}^2 T_{F1} q } \; \leq 1\; ,
\ee
and this bound must be satisfied in order to get a stable configuration. Moreover, from the boundary equation (\ref{boundary-eq}) we conclude that
\be 
l\geq \frac{q}{2}\left( \sqrt{1-\tilde{\b}^2} +1 \right)\label{bound}
\ee
and this bound defines the minimum value of quarks to create a stable baryon in the field theory. Furthermore, from the table \ref{table} we see 
that fundamental strings attached to D$2$-branes and D$4$-branes are possible just in the presence of magnetic field, therefore, the 
condition (\ref{bound}) also depends on the magnitude of the magnetic field on the brane worldvolume. We can easily see that 
for $\b=0$, that occurs when $M_{Dp}^L={L_{string}^2 T_{F1} q }$, the quark configurations are made of $q$-quarks, and the 
results reduce to what we obtained in the previous subsection.

Moreover, using (\ref{r-eq}) we can easily show that the size of this configuration is
\be 
l_{wb}=\frac{\tilde{u}^2}{u_0^3}\int_1^\infty \dd z \frac{1}{z^2\sqrt{z^4 - \frac{\tilde{u}^4}{u_0^4}}} \; ,
\ee
where $z=u/u_0$, and similarly to (\ref{size}), this result can be integrated in terms of hypergeometric functions, giving as there 
\be
l_{wb}=\frac{\tilde u^2}{u_0^3}{}_2F_1\left(\frac{1}{2},\frac{3}{4},\frac{7}{4};\frac{\tilde u^4}{u_0^4}\right).
\ee




\subsection{Giant gravitons: gravity calculation}

Giant gravitons in a gravitational solution correspond to solitonic states, obtained by wrapping D-branes on cycles in the geometry, which 
are therefore extended states, that are {\em moving in a direction transverse to the cycle},
and that are degenerate in energies with graviton states. A proper treatment of the giant gravitons will take a separate 
paper, so here we will just sketch the main points.

The GJV gravity solution with metric (\ref{metric}) is a solution of massive type IIA string theory, and we saw that there are wrapped D2-branes and D4-branes
that are tadpole-free (in the absence of magnetic flux), thus could be interpreted as as giant gravitons. Moreover, we have already said that we 
could obtain a nonzero mass (by a nonzero WZ term) for a moving brane, whereas the static brane will have vanishing mass. This is exactly the 
giant graviton case. 

1. We can obtain a standard kind of giant graviton by wrapping a D2-brane on an $S^2$. The $AdS_4$ metric to be considered is
\be 
\dd s^2_{AdS_4} = -(1+r^2) \dd t^2 + \frac{\dd r^2}{1 + r^2} + r^2 \dd \Omega^2_2 ,\label{ads4metric}
\ee
where $\dd \Omega^2_2= \dd \vartheta^2 + \sin^2 \vartheta \dd \varphi^2  $ parametrizes the two sphere $S^2$, on which the D2-brane is wrapped.

2. Another type of giant graviton is obtained by wrapping a D2-brane on an $\mathbb{CP}^1 =S^2$ inside the compact geometry. 
We can consider the  $S^2=\mathbb{CP}^1 \subset \mathbb{CP}^2$ defined by ${\cal J}$.
By considering the leading WZ term $\int C_3$ in the case when $\eta=\eta(t)$, we will obtain a nonzero contribution to the mass, 
proportional to 
\be
\frac{\sin^4\a }{3+\cos 2\a}\;\dot \eta\;,
\ee
but the DBI term for the moving brane will also be modified. 

3. Yet another type of giant graviton will be obtained by considering D4-branes wrapping the $\mathbb{CP}^2$. 
Again, the leading WZ term in $\int C5$  for $\eta=\eta(t)$ will give a nonzero contribution to the mass (note that $\tilde F_6=-e^{\phi/2}*\tilde F_4$), 

\subsubsection{D$2$-brane on $ S^2 \subset AdS_4 $}
 
We consider the metric (\ref{ads4metric}) for the $AdS_4$ space,
with the sphere $S^2$ parametrized by $(\vartheta,\varphi)$. The ansatz for the giant gravitons is given by
\be 
\xi^0=t\; , \xi^1=\vartheta\; , \xi^2= \varphi\; , \psi=\psi(t)\; ,
\ee
and we set $\lambda=\pi/2$ and $\a=\pi/2$. We study the giant gravitons dynamics following the well known recipe \cite{McGreevy:2000cw, Grisaru:2000zn, Hashimoto:2000zp}, in particular, giant gravitons in the GJV solution are similar to the configurations considered in the ABJM case \cite{Nishioka:2008ib, Hamilton:2009iv, Cardona:2014ora}. The induced metric on the brane is
\be 
\dd s^2_{D2}= L_{string}^2 \left[\left( \frac{9}{4} \dot{\psi}^2 -(1 + r^2) \right)\dd t^2 + r^2 \dd \Omega_2^2 \right]
\ee
and the DBI plus WZ action is
\be 
S_{D2}=-4\pi T_{D2} \frac{L_{string}^3}{g_s^\a} \int\dd t\; \left( r^2 \sqrt{1 + r^2 - \dot{\chi}^2 } -  r^3 \right)
\ee
where we have defined $\chi \equiv \frac{3}{2}\psi $. When we write the momentum $P_\chi$ as
\be 
P_\chi=\frac{4\pi T_{D2}L_{string}^3}{g_s^\a}\frac{r^2\dot{\chi}}{\sqrt{1+r^2-\dot{\chi}^2}}\equiv 
\frac{4\pi T_{D2}L_{string}^3}{g_s^\a}p\label{momentum}
\ee
and the Hamiltonian is simply
\be 
{\cal H}=\frac{4\pi T_{D2}L_{string}^3}{g_s^\a}\left( \sqrt{(1+r^2)(p^2+r^4)}-r^3 \right)
\ee
and this functional has two minima, with ${\cal H}=P_\chi$, namely $r=0,p$ and the latter we call giant graviton. Furthermore, if we insert the 
giant graviton into (\ref{momentum}) we find $\chi=t$, so that the angular velocity $\dot{\chi}$ equals the speed of light. 

As a technical detail, one may notice that the giant gravitons in \cite{Nishioka:2008ib, Hamilton:2009iv} move in a circle 
$S^1\subset \mathbb{CP}^3$ parametrized by the coordinate $\s$ in the Fubini-Study metric
\be 
\begin{split}
\dd s_{\mathbb{CP}^3}=& \dd \l^2 + \frac{1}{4}\sin^2 \l  (\dd \theta^2 + \sin^2 \theta \dd \phi^2) + \frac{1}{4}\cos^2 \l (\dd \tilde{\theta}^2 
+ \sin^2 \tilde{\theta} \dd \tilde{\phi}^2)\\
& + \frac{1}{4}\cos^2 \l \sin^2 \l (\dd \s + \cos \theta \dd \phi - \cos \tilde{\theta} \dd \tilde{\phi}) \; ,
\end{split}
\ee
that has a subspace ${\bb{CP}}^2$ for constant $(\tilde{\theta}, \tilde{\phi})$. Therefore, in order to have a proper identification of the giant 
graviton of this section with the giants of \cite{Nishioka:2008ib, Hamilton:2009iv}, we must consider the motion confined in the coordinate 
$\s$, instead of $\psi$, in squashed $6$-sphere of the GJV solution. The derivation is isomorphic to what we have done so far, so 
we will not repeat the calculations.

\subsubsection{D$2$-brane on the squashed $S^2 \subset {\cal S} S^5 $}

When we consider that the D$2$-branes wraps a two sphere inside the internal space ${\cal S} S^5$ defined by (\ref{int-space}), the calculations are 
more involved because of the non-diagonal terms in the metric, but the idea is essentially the same, with one remarkable difference, 
the role of the angular coordinates, as we will see below. 

The ansatz is given by
\be 
\xi^0=t\; , \xi^1=\theta\; , \xi^2= \phi\; , \psi=\psi(t)\; ,
\ee
and we set $r=0$ and constant  $\lambda$ and $\s$. The induced metric on the brane is
\be 
\begin{split}
\dd s^2_{D2} & = L_{string}^2 \left[(\Omega \dot{\psi}^2 -1 )\dd t^2 + \frac{\Xi \sin^2\l}{4}\dd \theta^2 + 
 2\Omega \frac{\dot{\psi} \sin^2\l \cos \theta}{2} \dd t \dd \phi \right.\\
& \left. + \frac{\sin^2\l}{4}\left[ \Xi(\sin^2 \theta + \cos^2\l\cos^2 \theta) + \Omega \sin^2\l \cos^2\theta \right] \dd \phi^2 
\right]\; ,
\end{split}
\ee
and the Dirac-Born-Infeld action reads
\begin{align}
 S_{DBI} & =-\frac{T_{D2} L_{string}^3 \pi}{2 g_s^\a}\int \dd \theta \dd t\; \sin^2 \l 
 \sqrt{\Xi
(\Xi\cos^2\theta \cos^2\l + \Xi \sin^2\theta + \Omega \cos^2\theta \sin^2\l)
}\; \times\nonumber \\
& \times
\sqrt{
1 - \frac{(\Xi\cos^2\theta \cos^2\l + \Xi \sin^2\theta -3 \Omega \cos^2\theta \sin^2\l)}{(\Xi\cos^2\theta \cos^2\l + \Xi \sin^2\theta 
+ \Omega \cos^2\theta \sin^2\l)}\Omega \dot{\psi}^2
}\; .
\end{align}
The $\theta$-integration can be carried out easily and gives an expression in terms of elliptic functions. But in order to to avoid unnecessary 
complications, we will keep it as it is. Together with the Wess-Zumino term
\be 
S_{WZ}=-\frac{T_{D2} L_{string}^3 \pi}{g_s^\a}\int \dd \theta\sin \theta \int \dd t\; \sin^2 \l \frac{3\sqrt{3} }{(3 + \cos 2 \a)^2}\sin^4 \a \dot{\psi}\; ,
\ee
we have the D$2$-brane action
\be 
S_{D2}=-\int \dd \theta \int \dd t \left( f(\a,\l;\theta) \sqrt{1 - g(\a, \l;\theta) \dot{\psi}^2} -h(\a,\l;\theta)\dot{\psi} \right)\label{B-action}
\ee
where we write the action in terms of the functions $f(\a,\l;\theta),\; g(\a,\l;\theta)$ and $h(\a;\theta)$ given by
\begin{align}
f(\a, \l;\theta) & = \frac{T_{D2} L_{string}^3 \pi}{2 g_s^\a} \sin^2 \l 
 \sqrt{\Xi
(\Xi\cos^2\theta \cos^2\l + \Xi \sin^2\theta + \Omega \cos^2\theta \sin^2\l)
}
\nonumber\\ 
& \equiv f_0(\a) \sin^2\l \sqrt{f_1(\a;\theta)\cos^2\l + f_2(\a; \theta)} \\
g(\a,\l;\theta) & = \frac{\Omega (\Xi\cos^2\theta \cos^2\l + \Xi \sin^2\theta -3 \Omega \cos^2\theta \sin^2\l)}{(\Xi\cos^2\theta \cos^2\l 
+ \Xi \sin^2\theta + \Omega \cos^2\theta \sin^2\l)}\nonumber\\
& \equiv \frac{g_1(\a;\theta)\cos^2\l + g_2(\a; \theta)}{f_1(\a;\theta)\cos^2\l + f_2(\a; \theta)}\\
h(\a,\l;\theta) & = - \frac{T_{D2} \pi L_{string}^3\sin \theta }{g_s^\a} \frac{3\sqrt{3}}{(3 + \cos 2 \a)^2}\sin^4 \a \sin^2\l 
\equiv h_1(\a;\theta)\sin^2\l \; .\label{h1}
\end{align}

For the sake of simplicity, we assume that $\l=\pi/2$.

With these definitions, the momentum (up to a $\theta$-integration) is
\be  
P_\psi= \frac{g f \dot{\psi}}{\sqrt{1 - g \dot{\psi}^2 }} + h\; ,\label{momentum2}
\ee
which we can invert to write $\dot{\psi}$ in terms of $P_\psi$, as
\be 
\dot{\psi} = \frac{ P_\psi - h}{\sqrt{(f g)^2 + g \left[P_\psi -h\right]^2}}\; .
\ee
The Hamiltonian becomes
\be 
\begin{split}
{\cal H} & = P_\psi \dot{\psi} - {\cal L}\\
& = \frac{1}{g}\sqrt{g(P_\psi-h)^2 + f^2 g^2}\; .\label{energy-giant}
\end{split}
\ee
Writing down explicitly the equation $\d_\a {\cal H}\equiv {\cal H}'=0$, we have 
\be 
-2g'{\cal H}^2 + \frac{1}{g}\left[ g' (2f^2 g+(P_\psi - h)^2) - 2 g (P_\psi - h)h' + 2g^2 f f'
\right]=0\; ,
\ee
and solving the condition for the angle $\a$ that follow from this equation, we find the pointlike and giant gravitons solutions.

The solution that corresponds to the pointlike graviton is given by $\sin \a\cos \a=0$. Since $\a=0,\pi$ are isolated conical singularities, 
the system is not well defined at these points. We set $\a=\pi/2$ for the pointlike gravitons. For the giant graviton solution, we must solve
the equation $\d_\a{\cal H}=0$, for $\a\neq \pi/2$, that gives an algebraic equation which must be solved for $X=\sin^2\a$.

The non-diagonal terms in the metric make the explicit verification of these algebraic conditions very involved. For simplicity then, we consider a dielectric 
brane, such that the electric field is conveniently chosen to cancel the non-diagonal terms of the metric \cite{Herrero:2011bk}. In this somewhat
contrived case, the D$2$-brane action (\ref{B-action}) reduces to
\be
 S_{D2} =- \int \dd \theta \int \dd t \left( f(\a, \l; \theta) \sqrt{1-\Omega \dot{\psi}^2} - h(\a,\l;\theta)\dot{\psi}
\right)\; ,
\ee
with momentum
\be 
P_\psi =  \frac{f \Omega \dot{\psi}}{\sqrt{1- \Omega \dot{\psi}^2}}+h \, \Rightarrow \, \dot{\psi}=\frac{P_\psi-h}{\sqrt{(f\Omega)^2 + \Omega (P_\psi-h)^2}}\; ,
\ee
and energy given by
\be 
{\cal H}=\frac{1}{\Omega}\sqrt{(f\Omega)^2 + \Omega (P_\psi-h)^2}\; ,
\ee
that is equivalent to the replacement $g \to \Omega$ in the Hamiltonian (\ref{energy-giant}). In this case the pointlike gravitons have again $\a=\pi/2$.

\subsubsection{D$4$-brane on the squashed $\mathbb{CP}^2 \subset {\cal S} S^5 $}

Finally, we consider a D$4$-brane wrapping the whole squashed complex projective space  $\mathbb{CP}^2$, so the induced metric on the D4-brane is
\be 
\begin{split}
\dd s^2_{D4} & =L^{2}_{string}\left\{ (-1 + \Omega \dot{\psi}^2)\dd t^2 + 2\frac{\Omega \dot{\psi}  \sin^2 \l}{2}(\dd \s 
+ \cos \theta \dd \phi)\dd t + \Xi \frac{\sin^2 \l}{4}\dd \theta^2 + \Xi \dd \l^2 \right.\\
& + \left.
\frac{\sin^2\l}{4}\left[
\Xi \sin^2 \theta \dd \phi^2 + (\Xi \cos^2 \l + \Omega \sin^2\l)(\dd \s + \cos\theta \dd \phi)^2
\right]
\right\}\; .
\end{split}
\ee

It gives the Dirac-Born-Infeld action
\be 
S_{DBI} =- \tilde{f}(\a)\int\dd t \dd \l \sin^3 \l \sqrt{\Omega (1 - \cos 2 \l) + \Xi (1- \Omega \dot{\psi}^2 )(1 + \cos 2 \l)} \;,
\ee
where
\be 
\tilde{f}(\a) = T_{D4} \frac{ \sqrt{2} L^5_{string}\pi^2\Xi^{3/2} }{g_s^\a}\; .
\ee

Given the $6$-form RR field $\widetilde{F}_6$ defined by $\widetilde{F}_6=-e^{\phi/2}\ast \widetilde{F}_4$, its restriction to ${\cal S}S^5$ is 
\be 
\left. \phantom{\frac{1}{1}} \widetilde{F}_6\right|_{{\cal S}S^5}=-\left(\frac{3L^3 e^{-\phi_0/4}}{4}\sqrt{\frac{3}{2}}\Delta 
\sqrt{\Omega}\Xi^2 e^{\phi/2}\sin^3\l \cos\l \sin \theta\right)\  \dd \a\dd \psi\dd \theta \dd \l \dd \phi \dd \s
\ee
where the wedge product of forms is implicit. We may write this expression as
\be 
\left. \phantom{\frac{1}{1}} \widetilde{F}_6\right|_{{\cal S}S^5} =  \dd C^{{\cal S}S^5}_5\; ,
\ee
where 
\be
 C^{{\cal S}S^5}_5= -\left(81 L^5 e^{\phi_0/4}\sqrt{\frac{3}{2}}
 A(\a) \sin^3\l \cos\l \sin \theta\right)\  
 \dd \psi\dd \theta \dd \l \dd \phi \dd \s
\ee
and
\be 
A(\a)=\int \dd\a \frac{\sin^5 \a}{( 3 + \cos 2 \a )^2}= \frac{1}{2}\arctan (\cos \a) - \frac{\cos \a}{4} - \frac{\cos \a}{3 + \cos 2 \a}\; .
\ee

Therefore, the Wess-Zumino term for a D$4$-brane wrapping the squashed space $\mathbb{CP}^2$ is
\be 
S_{WZ}= - T_{D4}\int C_{5}^{{\cal S}S^5} = \tilde{g}(\a) \int \dd t \dd \l \sin^3\l \cos \l \dot{\psi}\; ,
\ee
where we have defined $\tilde{g}(\a)=324 T_{D4} \sqrt{6} \pi^2 e^{\phi_0/4}A(\a)$. Finally, the D$4$-brane action is simply 
\be 
\begin{split}
S_{D4} & =-\int\dd t \dd \l 
\left( \tilde{f}(\a) \sin^3 \l \sqrt{\Omega (1 - \cos 2 \l) + \Xi (1- \Omega \dot{\psi}^2 )(1 + \cos 2 \l)}  \right.\\
&\hspace{4.0cm} \left. 
- \tilde{g}(\a) \sin^3\l \cos \l \dot{\psi}
\right)\; .
\end{split}
\ee
We leave the $\l$-integral untouched, since the result is quite complicated. Therefore, the Hamiltonian (up to a $\l$-integration) is
\be 
\begin{split}
{\cal H}= & \frac{\sqrt{\Xi (1 + \cos 2 \l) + \Omega  (1 - \cos 2 \l)}}{\Xi \Omega  (1 + \cos 2 \l)}\times \\
& \times\sqrt{\tilde{f}^2 \sin^2\l \Xi^2 \Omega^2  (1 + \cos 2 \l)^2 + \Xi \Omega (P_\psi - \tilde{g} \sin^3\l \cos \l)^2}\; ,
\end{split}
\ee
and again, studying the condition that follows from $\d_\a {\cal H}=0$ we find the pointlike and giant graviton solutions.

\subsection{Giant gravitons: field theory operators}

As usual, the field theory operators corresponding to the giant gravitons are subdeterminants, of the type
\be
\epsilon_{a_1...a_k a_{k+1}...a_N}\epsilon^{b_1...b_k a_{k+1}...a_N}({Z_{b_1}}^{a_1}....{Z_{b_k}}^{a_k})\;,
\ee
or with $W$ and $T$ replacing any of the $Z$ fields. More generally, the operators are Schur polynomials for some complicated 
reprentation of the symmetry group. 

-For the D2-brane wrapping $S^2\subset \mathbb{CP}^2$, the fields inside the bracket are $Z$'s and $W$'s in an $SU(2)$ invariant combination, 
forming representations of $SU(2)$. 

-For the D2-brane wrapping $S^2\subset AdS_4$, we should replace the $Z$'s with $D_a Z$'s, where $a$ corresponds to the two angles on 
$\Omega_2$, the spatial 2-sphere in $AdS_4$. Again, the combinations of $D_aZ$'s should be in some representation of $SU(2)$.

-For the D4-brane wrapping $\mathbb{CP}^2$, we should have all the $Z,W,T$ inside the bracket, transforming under $SU(3)$ (the symmetry
group of $\mathbb{CP}^2$), i.e. forming a representation of this symmetry group.

\subsection{Gauge coupling from wrapped brane}

One can easily show that given a induced metric on a D$p$-brane wrapping an $n$-cycle $\Sigma^n$ of the form,
\be 
\dd s^2_{Dp}=e^{2A}\eta_{\m\n} \dd x^\m \dd x^\n  + \dd s^2_{\Sigma}\; ,
\ee
the gauge coupling defined on the Minkowski worldvolume of the brane is
\be 
\frac{1}{g_{YM}^2}= T_{Dp} (2\pi \a')^2 \int_{\Sigma} \dd^n \xi e^{(3+n-p)A-\phi}\sqrt{\det\left( G + B \right)_{ab}}
\ee
where $\phi$ is the dilaton and $G_{ab}$ and $B_{ab}$ are the metric and the Kalb-Ramond field along the $n$-cycle $\Sigma^n$.

In the GJV solution, we can wrap a probe D$4$-brane on a $2$-cycle parametrized by $(\theta, \phi)$ at $\lambda=0$. 
In this case, the (3 dimensional) gauge coupling of the probe is given by
\be 
\frac{1}{g_{YM}^2}=\frac{\Delta^{3/2} \Xi e^{-\phi/4}}{4\pi \a'^{3/2}}\frac{1}{y}\sim \frac{ \left(N^2 k \right)^{1/3}}{y}=\frac{N}{\lambda^{1/3}}\frac{1}{y}\; ,
\ee 
where we have used the $AdS_4$ coordinate system $\dd s^2_{AdS4}=\left(\dd y^2 + \eta_{\m\n} \dd x^\m \dd x^\n \right)/y^2$.

We observe that, as expected, the probe gauge coupling is now given by the curvature in string units as $\propto \lambda^{1/6}$.


\section{Strings in the geometry}

\subsection{Wilson loops}

Despite the fact that the Yang-Mills gauge coupling is dimensionful in three dimensions, and that we are studying the gravity dual of a conformal 
field theory, Wilson loops are interesting observables to consider. Using the standard prescription of \cite{Maldacena:1998im, Sonnenschein:1999if, Nunez:2009da} in the metric background (\ref{metric}), we consider a string configuration that consists of an open string with its endpoints attached to the $AdS_4$ boundary, $u\to \infty$, at the points $x=0$ and $x=L_{q\bar{q}}$, and that extends into the bulk  and reaches a minimum radial coordinate $u_0$ exactly at $x=L_{q\bar{q}}/2$.

The ansatz for this configuration is given by $t=\tau$, $x=x(\s)$ and $u=u(\s)$, so that the Nambu-Goto action is
\be 
S=\frac{L_{string}^2}{2\pi \a'}\int \dd \tau \dd \s \sqrt{ u^4 x'^2+ u'^2}
\ee
where the prime denotes derivative with relation to $\s$. From the equations of motion for this system we have
\begin{align}
\frac{\dd u}{\dd x} & =- \frac{u^2}{u_0^2}\sqrt{u^4-u_0^4}\quad \text{for} \quad x<\frac{L_{q\bar{q}}}{2}\\
\frac{\dd u}{\dd x} & =+ \frac{u^2}{u_0^2}\sqrt{u^4-u_0^4}\quad \text{for} \quad x>\frac{L_{q\bar{q}}}{2}\; .
\end{align}
Using these equations, we can determined that the quark-antiquark distance is given by
\be 
L_{q\bar{q}}=\frac{2}{u_0}\int_{1}^\infty \dd y \frac{1}{y^2\sqrt{y^4-1}}=\frac{1}{u_0}\frac{(2\pi)^{3/2}}{\Gamma(1/4)^2}\; .
\ee
where $y=u/u_0$. 

Furthermore, if we replace this solution, the quark-antiquark potential diverges, so that we renormalize this system by removing the mass of two $W$-bosons, that consists of two strings stretching from the boundary into the bulk of the AdS space. Therefore, we find that the potential is given by
\be 
E_{q\bar{q}}=-\frac{(2\pi)^2}{\Gamma(1/4)^4}\frac{L_{string}^2}{\a' L_{q\bar{q}} }\; ,
\ee
which is similar to the result of \cite{Maldacena:1998im}. It still has the conformal form $\propto 1/L_{q\bar q}$, 
but there is an interesting difference when we write it in terms of the 
't Hooft coupling $\lambda$. In the $AdS_5 \times S^5$ case  one obtains a quark-antiquark potential that behaves as the square 
root of the 't Hooft coupling, $E_{q\bar{q}}\sim\sqrt{\lambda}/L_{q\bar{q}}$, but in our case, 
since the 't Hooft coupling is $\lambda= N/k$ (see also \cite{Aharony:2010af, Gaiotto:2009mv}), we have 
the one third power of the coupling,
\be 
E_{q\bar{q}}=-\frac{2^{13/6}\pi^3}{3^{2/3}\Gamma(1/4)^4}\frac{\sqrt{5+\cos 2\a}\lambda^{1/3}}{L_{q\bar q}}
\sim\frac{\lambda^{1/3}}{L_{q\bar{q}}}.
\ee

\subsection{Operators of large spin}

Strings rotating around the circle $S^2 \in AdS_4$ in global coordinates will be dual to high spin operators \cite{Gubser:2002tv}, 
and  the anomalous dimension of these operators in SYM  is generically
\be 
E-S= f(\lambda) \ln S\; .
\ee

We take the $AdS_4$  metric as
\be 
\dd s^2_{AdS_4}=-\cosh^2 \r \dd t + \dd \r^2 + \sinh^2 \r (\dd \vartheta^2 + \sin^2 \vartheta \dd \varphi^2)
\ee
and we consider that the string is at the equator of $S^2$, $\vartheta=\pi/2$ and
\be 
t=\tau\; , \; \r=\r(\s)\; ,\;  \varphi=\omega \tau\; .
\ee
The Nambu-Goto action is then
\be 
S_{NG}=-\frac{4L_{string}^2}{2\pi \a'}\int_0^{\r_0} \dd \r \sqrt{\cosh^2 \r - \sinh^2\r \dot{\varphi}^2}
\ee
and the $4$ comes from the fact that the closed string forms a loop that consists of two parallel segments, and spins as a rigid rod around its center at $\r=0$ and extends to $\r_0$, determined by $\omega^2=\coth^2\r_0$ \cite{Gubser:2002tv}. Similarly, the string energy and spin are
\begin{align}
 E& = \frac{4L_{string}^2}{2\pi \a'}\int_0^{\r_0}\dd \r \frac{\cosh^2\r}{\sqrt{\cosh^2 \r - \sinh^2\r \omega^2}}\\
 S& = \frac{4L_{string}^2}{2\pi \a'}\int_0^{\r_0}\dd \r \frac{\omega \sinh^2\r}{\sqrt{\cosh^2 \r - \sinh^2\r \omega^2}} \; ,
\end{align}
respectively.

Consider now that the string length is much larger than the $AdS_4$ radius, $L\sim L_{string}$, then $S>>L_{string}^2$, and it corresponds to the limit $\omega\to 1$. Taking $\omega\approx 1 + 2\eta$\; for $\eta<<1$, one can easily obtain $2\r_0=\ln (1/\eta)$ and 
\be
 E = \frac{L_{string}^2}{2\pi \a'}\left( 1/\eta + \ln (1/\eta)+\cdots\right)\; , \quad S = \frac{L_{string}^2}{2\pi \a'}\left( 1/\eta - \ln (1/\eta)+\cdots\right)  \; ,
\ee
so that the cusp anomalous dimension is found to be
\be
f(\lambda)= L^2_{string}/ \pi\a'\propto \l^{1/3}\;,
\ee
and the proportionality constants are easily obtained from (\ref{quant}). Note that, in contrast to \cite{Gubser:2002tv, Aharony:2008ug},  the anomalous cusp dimensional scales as $\l^{1/3}$ instead of $\lambda^{1/2}$.


\section{Entanglement entropy}

The metric of the GJV solution (\ref{metric}) has the form of a warped, squashed $AdS_4\times S^6$. Because of this, it is worth to 
consider the calculation of the holographic entanglement entropy, acording to the prescription of Ryu and Takayanagi \cite{Ryu:2006bv}, 
extended by Klebanov, Kutasov and Murugan to more general gravity duals in \cite{Klebanov:2007ws}.

We consider a $d+2$ dimensional gravity dual, with a $d+1$ dimensional quantum field theory at its boundary, and we want to compute the entanglement 
resulting from introducing a closed $d-1$ dimensional spatial
surface $\d A$ that separates the spatial region $A$ from the rest of $d$ dimensional space (at constant time). 
Consider then the $d$ dimensional spatial surface $\gamma$ in the gravity dual that ends on the same boundary surface, $\d \gamma=\d A$. 
Then the holographic entanglement entropy is given by 
\be
S_A=\frac{1}{4 G_N^{(d+2)}}\int_\gamma d^d \sigma \sqrt{G^{(d)}_{\rm ind}}.
\ee
For a 10 dimensional metric, moreover in the presence of a nontrivial dilaton field $\phi$, we have instead
\be
S_A=\frac{1}{4 G_N^{(10)}}\int_\gamma d^8\sigma e^{-2\phi}\sqrt{G^{(8)}_{\rm ind}}.
\ee
Here $G_{\rm ind}$ is the induced string frame metric on the surface $\gamma$. 

The string frame metric $ds^2_{\rm string}=e^{\phi/2}ds^2_{\rm E}$ for the GJV solution, (\ref{metric}), is 
\be
ds^2_{\rm string}=L^2 e^{\phi_0/2}(5+\cos\a)^{1/2}\left[ds^2_{(AdS_4)}+\frac{3}{2}d\a^2+\Xi ds^2_{\mathbb{CP}^2}+\Omega \eta^2\right]\;,
\ee
which can be put in the form
\bea
ds^2_{\rm string}&=&A(z)[ds^2_{(AdS_4)}+dz^2]+g_{ij}(z)d\theta^i d\theta^j\cr
&=& A(z)\left[\frac{dy^2-dt^2+dx_1^2+dx_2^2}{y^2}+dz^2\right]+g_{ij}(z,...)d\theta^id\theta^j\;,
\eea
where $z=\a\sqrt{3/2}$ and $A(z)=L^2 e^{\phi_0/2}(5+\cos \a)^{1/2}$.

This type of metric is somewhat more general than metrics considered before, for instance than in \cite{Klebanov:2007ws}. 

Consider the region $A$ to be the interval $(-l,l)$ in one of the spatial directions, called $x$, times the full space in the others, so that $\d A$ are the two 
endpoins of the interval, times the full space in the other directions. The surface $\gamma$ in the gravity dual is then a curve $y(x)$ with boundary 
conditions $y(-l)=0=y(l)$, where $y$ is the radial coordinate of AdS. This is the same as in the simple $AdS_{d+2}/CFT_{d+1}$ case.
In our $AdS_4$ case, $d=2$, and we can take $x_1=x$, $x_2$ being trivial, but the discussion is more general than this $d=2$ case. 

The induced metric on $\gamma$ (with $y=y(x)$) is then in this more general case (with $d-1$ coordinates $\vec{x}$ and one $x$)
\be
ds^2_{\rm string}=A(z)\left[\frac{dx^2(1+y'^2)+d\vec{x}^2}{y^2}+dz^2\right]+g_{ij}(z,...)d\theta^i d\theta^j.
\ee
Denoting 
\be
V_{\rm int}(z)\equiv d^{D-d-3}\theta\int \sqrt{\det g_{ij}}\;,
\ee
the entanglement entropy of the interval becomes
\bea
S_A&=&\frac{1}{4G_N^{(10)}}\int d^{d-1}\vec{x} \int dz \int_{-l}^l dx e^{-2\phi}V_{\rm int}(z) y(x)^{-d}A(z)^{\frac{d+1}{2}}\cr
&=&\left\{\frac{1}{4G_N^{(10)}}\left[\int dz e^{-2\phi(z)}V_{\rm int}(z)A(z)^{\frac{d+1}{2}}\right]\right\}\int d^{d-1}\vec{x}\int_{-l}^{+l}dx\; y(x)^{-d}\sqrt{1+
(y'(x))^2}.\cr
&&
\eea

The square brackets offer a constant renormalization of $G_N^{(10)}$, but otherwise the result is the same as for $AdS_{d+2}=AdS_4$.
The same profile for $y(x)$ is obtained, and the same final result for the entanglement entropy as a function of $l$.

\section{Giant magnons}

We now turn to some large string theory objects, specifically giant magnons, that come from long (macroscopic) strings in the gravity dual, 
and correspond in the gauge theory to spin chain states of large momentum.

\subsection{Gravity calculation}

\subsubsection{Review of $AdS_5\times S^5$ giant magnons}

The giant magnons were defined by Hofman and Maldacena in \cite{Hofman:2006xt}.
Since the regular magnons (spin excitations) correspond to quantum states of the string on the pp wave, obtained as a Penrose limit of the 
gravity dual, giant magnons corespond to classical, i.e. long, strings in the gravity dual. Specifically, since we look for giant magnons in the 
$SU(2)=SO(3)$ sector of the field theory (made up from two complex scalars $Z$ and $W$), in the gravity dual the motion of the strings is 
restricted to $\mathbb{R}_t\times S^2$. The spatial $S^2$ is embedded in the $S^5$ of $AdS_5\times S^5$ as
\be
ds^2_{S^5}=\sin^2\theta d\phi^2 +d\theta^2+\cos^2\theta d\Omega_3^2\;,
\ee
by $d\Omega_3=0$. We consider the gauge $t=\tau$, $\phi-t=\phi'\equiv \sigma$, where $(\tau,\sigma)$ are the string worldsheet coordinates. 
Then the induced metric on the string worldsheet is 
\bea
ds^2_{\rm induced}&=&-d\tau^2-\frac{ds^2_{S^5}}{d\tau^2}d\tau^2+\frac{ds^2_{S^5}}{d\sigma^2}d\sigma^2+2d\sigma d\tau \frac{ds^2_{S^5}}{d\tau\; d\sigma}\cr
&=&-\cos^2\theta d\tau^2+(\sin^2\theta+\theta'^2)d\sigma^2+2\sin^2\theta d\sigma\; d\tau.
\eea
Then the string action is, for $R^2/(2\pi \a')=\sqrt{\lambda}$,
\bea
S&=&\frac{\sqrt{\lambda}}{2\pi}\int d\sigma \; d\tau\sqrt{-\det g_{ab,{\rm induced}}}\cr
&=&\frac{\sqrt{\lambda}}{2\pi}\int dt \int d\phi'\sqrt{\cos^2\theta \theta'^2+\sin^2\theta}\;,
\eea
and its (integrated) equation of motion is 
\be
\sin \theta=\frac{\sin \theta_0}{\cos\phi'}\;,
\ee
and $\theta_0$ is an integration constant, such that $\phi'\in [-(\pi/2-\theta_0),+(\pi/2-\theta_0)]$. This is a string that rotates around $S^2\in S^5$, while 
describing an arc that starts and ends on the Equator, with an angle $\Delta \phi=\Delta\phi'=2(\pi/2-\theta_0)$ between the enpoints. 

The (lightcone) energy of the string is 
\be
\epsilon = E-J =\frac{\sqrt{\lambda}}{\pi}\left|\sin\frac{p}{2}\right|.
\ee

\subsubsection{Giant magnons in the gravity dual}

In the GJV gravity dual, we could consider various giant magnon solutions (spinning classical strings). But here we are interested simply in 
embedding the simple $SU(2)$-invariant giant magnons from the $AdS_5\times S^5$ case in the present context. For that, it suffices to 
find an $S^2$ fiber inside the metric (\ref{metric}). 

We can write the $\mathbb{CP}^2$ metric as (see (\ref{cp2}))
\be
ds^2_{\mathbb{CP}^2}=d\lambda^2+\frac{1}{4}\sin^2\lambda (\Sigma_1^2+\Sigma_2^2+\cos^2\l \Sigma_3^2)\;,
\ee
where the left-invariant one-forms are 
\bea
\Sigma_1&=& -\sin \sigma d\theta +\cos\sigma\sin\theta d\phi\cr
\Sigma_2&=& \cos\sigma d\theta +\sin\sigma \sin \theta d\phi\cr
\Sigma_3&=& d\sigma+\cos\theta d \phi.
\eea
Here $\lambda\in [0,\pi/2]$, $\sigma\in [0,4\pi]$, $\theta\in [0,\pi]$ and $\phi\in [0,2\pi]$. 
Note that 
\be
\Sigma_1^2+\Sigma_2^2=d\theta^2+\sin^2\theta d\phi^2\;,
\ee
so this combination forms an $S^2$. 

We then have two obvious possibilities for $S^2$ embedding:

-$\lambda=\pi/2$, $\theta=\theta_0$, which gives (the range for $\theta$ and $\phi$ is also correct)
\be
ds^2_{\mathbb{CP}^2}\rightarrow \frac{1}{4}(d\theta^2+\sin^2\theta d\phi^2)=\frac{1}{4}ds^2_{S^2}.\label{s2embed1}
\ee

-$\sigma=\sigma_0$, $\phi=\phi_0$, which gives
\be
ds^2_{\mathbb{CP}^2}\rightarrow (d\lambda^2+\sin^2\lambda(d(\theta/2))^2\;,
\ee
however with the range $\theta/2\in [0,\pi/2]$, $\lambda\in[0,\pi/2]$. Since for the giant magnon solution we would need $\theta/2$ to be in $[0,2\pi]$, 
we could extend the domain by symmetry (the metric is unchanged for the extension).

In the gravity dual, the action will be proportional to 
\be
\frac{L^2_{\rm string}}{2\pi \a'}\propto \left(\frac{N}{k}\right)^{1/3}=\sqrt{\lambda^{2/3}}\;,
\ee
so we would obtain a giant magnon energy of 
\be
\epsilon=E-J\propto \sqrt{\lambda^{2/3}}\left|\sin\frac{p}{2}\right|.\label{lambda23}
\ee

\subsection{Field theory}

As we saw, the dual field theory has 3 complex scalars in chiral multiplets, transforming under $SU(3)$, $Z,W,T$. For the giant magnons, we 
want to consider an $SU(2)$ (=$SO(3)$) sector, that corresponds to the motion on the $S^2$ with the same symmetry. We choose $Z$
and another scalar $\Psi$ to be 
the fields transforming under it. 

We consider a spin chain formed from $Z$ and $\Psi$ in the "dilute gas approximation", where the states are formed mostly from 
$Z$'s, with just a few $W$ excitations, similar to the BMN construction in 3+1 dimensions \cite{Berenstein:2002jq}. As there, the 
ground state is 
\be
|0,p^+\rangle=\frac{1}{\sqrt{J}N^{J/2}}\Tr[Z^J].
\ee
We consider a few excitations made by inserting in the $l$'th position $\Psi e^{\frac{2\pi i n l}{J}}$ inside the trace. Considering 
$\{W,T\}=\{\Phi^m\},$ $m=1,2$, we obtain a priori different spin chains for the $\Psi$ insertion being $\Phi^m, \bar \Phi^m$ or $\bar Z$. 

Similarly to the case of 3+1 dimensional ${\cal N}=4$ SYM, described in Appendix A of \cite{Cardona:2014ora}, but more similar to the ABJM case
of the same paper,
we obtain a Hamiltonian with eigenenergies of 
\be
\epsilon(p)=\sqrt{1+\frac{f_a(\lambda)}{\pi^2}\sin^2\frac{p}{2}}\;,
\ee
where $p=2\pi n/J$ is the momentum of the giant magnon, and $f_a(\lambda)$ is a function that changes its form depending on the 
$\Psi$ insertion, $(\Phi^m,\bar\Phi^m,\bar Z)$. For a regular magnon with $n\sim 1$, and in the limit $\lambda/J^2$ = finite 
(as $\lambda,J\rightarrow \infty$) we would have 
\be
\epsilon(p)\sim \sqrt{1+\frac{f_a(\lambda) n^2}{J^2}}\;,
\ee
which is the result in the BMN limit, corresponding to the Penrose limit. But if we take instead the limit $\lambda $ fixed and large, $p\sim 1$, 
we get 
\be
\epsilon(p)=\frac{\sqrt{f_a(\lambda)}}{\pi}\left|\sin\frac{p}{2}\right|\;,
\ee
and from the giant magnon solution in the gravity dual (\ref{lambda23}), we expect $f_a(\lambda)\propto \lambda^{2/3}$. 

\subsubsection{Hamiltonian and diagonalization}

We reproduce some of the steps in Appendix A of \cite{Cardona:2014ora},  that give the energies above. 
The introduction of the various fields inside the trace (which gives a cyclic ordering) can be described by the introduction of Cuntz oscillators $a_i$, 
$i=1,...,J$ acting on a vacuum, i.e. something like $a^\dagger...a^\dagger b^\dagger...a^\dagger|0\rangle$, where $a$ and $b$ correspond to $Z$ and 
$W$, respectively.  The Cuntz oscillators satisfy
\be
a_ia^\dagger_j=\delta_{ij};\;\;\;\sum_i a^\dagger_i a_i=1-|0\rangle\langle 0|;\;\;\; a_i|0\rangle=0.
\ee
If we only have a few "impurities" $W$ inside the trace, we can switch to a description in terms of {\em independent} Cuntz oscillators 
$b^\dagger_j$ at each site, satisfying 
\be
b_ib_i^\dagger=1;\;\;\; b^\dagger_i b_i=1-(|0\rangle\langle 0|)_i;\;\;\; b_i |0\rangle_i=0\;,
\ee
and 
\be
[b_i,b_j]=[b^\dagger_i,b_j]=[b^\dagger_i,b^\dagger_j]=0,\;\;\;\forall i\neq j.
\ee
Defining the Fourier modes $b_n$ by
\be
b_j=\frac{1}{\sqrt{J}}\sum_{n=1}^J e^{\frac{2\pi i j n}{J}}b_n\;,
\ee
and acting only on states in the dilute gas approximation, i.e. 
\be
|\psi_{\{n _i\}}\rangle=|0\rangle_1...|n_{i_1}\rangle_{i_1}...|n_{i_k}\rangle_{i_k}...|0\rangle_J\;,
\ee
one obtains the commutation relations
\be
[b_n,b_m^\dagger]|\psi_{\{n _i\}}\rangle=\left(\delta_{nm}-\frac{1}{J}\sum_k e^{\frac{2\pi i i_k(m-n)}{J}}\right)|\psi_{\{n _i\}}\rangle;\;\;\;
[b_n,b_m]=[b_n^\dagger,b_m^\dagger]=0\;,
\ee
so, up to $1/J$ corrections, the usual oscillator commutation relations. 

The interaction term in the Lagrangean in 2+1 dimensions is
\be
L_{\rm int}=-\frac{4\pi^2}{k^2}\Tr\left([[\bar \Phi^i,\bar\Phi^j,\bar \Phi^k][[\bar \Phi^j,\Phi^j], \Phi^k]\right)\;,
\ee
but where  in the IR we must impose $[\Phi^i,\Phi^j]=0$ from the superpotential term, and it
becomes equivalent, under the definition of the discretized field $\phi_j=(b_j+b_j^\dagger)/\sqrt{2}$, to ($\lambda=N/k$) 
\be
-2\lambda^2\sum_j(\phi_j-\phi_{j+1})^2.
\ee
Then the total Hamiltonian acting on the states created by Cuntz oscillators, including a free part, is found to be 
\be
H=\sum_{j=1}^J\frac{b_jb^\dagger_j+b^\dagger_j b_j}{2}+\lambda^2\sum_{j=1}^J(b_{j+1}+b_{j+1}^\dagger)(b_j+b_j^\dagger)-
(b_j+b_j^\dagger)^2\;,
\ee
where $\lambda=g^2_{YM}N$. This Hamiltonian can be diagonalized by the redefinition 
\be
b_{n}=\frac{c_{n,1}+c_{n,2}}{\sqrt{2}};\;\;\;\;
b_{J-n}=\frac{c_{n,1}-c_{n,2}}{\sqrt{2}}\;,
\ee
followed by the Bogoliubov transformation
\bea
&& \tilde{c}_{n,1}=a_nc_{n_1}+b_nc_{n,1}^\dagger\nonumber\\
&&\tilde{c}_{n,2}=a_nc_{n_1}-b_nc_{n,1}^\dagger\nonumber\\
&&a_n=\frac{(1+\alpha_n)^{1/4}+(1+\alpha_n)^{-1/4}}{2}\nonumber\\
&&b_n=\frac{(1+\alpha_n)^{1/4}-(1+\alpha_n)^{-1/4}}{2}\cr
\alpha_n&=&\lambda^2 (\cos (2\pi n/J)-1)= -2\lambda^2 \sin ^2\frac{\pi n}{J}  \;,
\eea
leading finally to 
\be
H=\sum_{n=1}^{J/2}\omega_n\left[\frac{\tilde{c}_{n,1}^{\dagger}\tilde{c}_{n,1}+
\tilde{c}_{n,1}\tilde{c}_{n,1}^{\dagger}}{
2}+\frac{\tilde{c}_{n,2}^{\dagger}\tilde{c}_{n,2}+\tilde{c}_{n,2}\tilde{
c}_{n,2}^{\dagger}}{2}\right]\;,
\label{hami}
\ee
where 
\be
\omega_n=\sqrt{1+4|\alpha_n|}=\sqrt{1+ 8\lambda^2 \sin^2 \frac{\pi n}{J}}.
\ee

In this section we have studied giant magnons, but an obvious question is what happens with regular magnons. They will give the usual pp wave (Penrose) limit of the gravity dual. This limit will be studied further in \cite{soon}.

\section{Conclusions}

In this paper we have studied various observables for the GJV duality between the warped, squashed $AdS_4\times S^6$ background and 
the 3 dimensional ${\cal N}=2$ SYM-CS theory at level $k$ at the fixed point. 

We have found that static wrapped D-branes give baryon vertices. At zero magnetic flux, we have D0-brane and D6-brane baryons, as well as 
their bound states, and they present evidence for the existence of a level-rank duality: besides the baryon vertex for $N$ quarks we also have a 
baryon vertex for $k$ quarks, and their masses are proportional to $N$ and $k$, respectively, other than the $\lambda^{1/6}$ piece. 
At nonzero magnetic flux, we have some interesting new baryon vertices that depend on the flux ${\cal N}$. Brane probes give a gauge coupling 
proportional to $\lambda^{1/6}$. 

Moving wrapped D-branes give giant gravitons, and we have found that we can have a D2-brane wrapped on $\mathbb{CP}^1$ or a D4-brane 
wrapped on $\mathbb{CP}^2$, both moving in the transverse $\eta$ (or $\psi$) direction. They correspond as usual to subdeterminant (or more 
generally Schur polynomial) operators. 

Entanglement entropy, calculated from the holographic prescription generalizing the one of Ryu and Takayanagi, gives nothing new with 
respect to AdS space, other than a renormalization of the Newton's constant. Wilson loops and the anomalous dimensions of operators of large 
spin are obtained from long strings in the dual, and both are proportional to $\lambda^{1/3}$.

Finally, giant magnons are found, also long strings in an $S^2$ subspace of the geometry, that correspond to spin chain operators in the 
field theory. Their Hamiltonian gives eigenergies consistent with the string calculation, and similar to the 4 dimensional 
SYM and ABJM cases. 

The GJV duality has many more important issues to be studied. A novel one is the level-rank duality discovered here. Others, like the 
Penrose limit, will be studied elsewhere \cite{soon}. It seems like this is just the tip of an iceberg, that should see many new developments in the near 
future. 

{\bf Acknowledgments} 
The work of HN is supported in part by CNPq grant 301219/2010-9 and FAPESP grant 2014/18634-9. The work of TA is supported by the \emph{Rec\'em Doutor} grant from PROPe-UNESP. We would like to thank Georgios Itsios and \'{E}oin \'{O} Colg\'{a}in for comments.

\appendix

\section{Conventions}\label{conventions}

The metric of the space $AdS_4$, normalized as $R^{AdS_4}_{mn}=-3g_{mn}^{AdS_4}$, is
\begin{align}
\dd s^2_{AdS_4} &= \frac{\dd u^2}{u^2}+u^2 \dd x^\m \dd x_\m\\
&= \frac{1}{y^2}\left( \dd y^2+ \dd x^\m \dd x_\m \right)\\
&= -(1+r^2) \dd t^2 + \frac{\dd r^2}{1 + r^2} + r^2 \dd \Omega^2_2 \; .
\end{align}
or in global coordinates
\be 
\dd s_{AdS_4}^2=- \cosh^2\rho \dd t^2 + \dd \rho^2 + \sinh^2 \rho \dd \Omega^2_2 \; ,\label{ads4}
\ee
where $\dd \Omega^2_2 = \dd \vartheta^2 + \sin^2 \vartheta \dd \varphi^2$ with $\vartheta\in [0,\pi]$ and $\varphi\in [0,2\pi]$, defines a two sphere $S^2$.

In addition, the metric of the complex projective space $\mathbb{CP}^2$, normalized in such a way that $R^{\mathbb{CP}^2}_{mn}=6g_{mn}^{\mathbb{CP}^2}$, see \cite{Pope:1980ub, Cardona:2014ora}, is
\begin{align}
 \dd s^2_{\mathbb{CP}^2}&=\dd\l^2+\frac{1}{4}\sin^2\l\left\{\dd\theta^2+\sin^2\theta \dd\phi^2+\cos^2\l(\dd\sigma+\cos\theta \dd\phi)^2\right\}\\
 &=\dd\l^2+\frac{1}{4}\sin^2\l\left\{\Sigma_1^2+\Sigma^2_2+\cos^2\l\;\Sigma_3^2\right\}\label{cp2}
\end{align}
where $0\leq 2\l, \theta, \phi/2 , \s/4\leq \pi $. In this coordinate system, the K\"ahler potential can be written as
\be 
\omega=\frac{\sin^2\l}{2}(\dd \s + \cos \theta \dd\phi)\; ,
\ee
with K\"ahler-form ${\cal J} =\frac{1}{2}\dd \omega$ and volume form $vol(\mathbb{CP}^2)=\frac{i^2}{2}{\cal J}\wedge{\cal J}$ and the volume of $\mathbb{CP}^2$ is $\int vol(\mathbb{CP}^2)=\frac{\pi^2}{2}$. Since, for fixed $\s$ and $\l=\pi/2$ we have a $\mathbb{CP}^1$ then
\be 
\int_{\mathbb{CP}^1} {\cal J}=-\pi \qquad \text{and}\qquad \int_{\mathbb{CP}^2} {\cal J} \wedge {\cal J}= -\pi^2\; .
\ee

Finally, it will be useful to consider the $SU(2)$ symmetry of this metric. Then, we can write the metric in terms of the Maurer-Cartan forms of the group $SU(2)$ given by
\begin{align}
L_1&=\frac{1}{\sqrt{2}}(-\sin\s d\theta+\cos\s\sin\theta d\phi)\equiv \frac{1}{\sqrt{2}}\Sigma_1\nonumber\\
L_2&=\frac{1}{\sqrt{2}}(\cos\s d\theta+\sin\s\sin\theta d\phi)\equiv \frac{1}{\sqrt{2}}\Sigma_2\\
L_3&=\frac{1}{\sqrt{2}}(d\s+\cos\theta d\phi)\equiv \frac{1}{\sqrt{2}}\Sigma_3.\nonumber
\end{align}
that satisfies $\dd L^i=\frac{\epsilon^{ijk}}{\sqrt{2}}L^j\wedge L^k$ and $\Sigma_1 \wedge \Sigma_2 \wedge \Sigma_3= \sin \theta \dd \theta\wedge \dd \sigma \wedge \dd \phi$.

\bibliographystyle{utphys}
\bibliography{GJVrefs}{}

 
\end{document}